\documentclass[11pt,a4paper]{article}
\usepackage{jheppub}
\usepackage{amsmath}
\usepackage{amssymb}
\usepackage{float}
\usepackage{hyperref}
\begin{document}

\numberwithin{equation}{section}
\numberwithin{table}{section}
\newtheorem{example}{Example}[section]
\newtheorem{conjecture}{Conjecture}[section]

\title{Hori-Vafa mirror periods, Picard-Fuchs equations, and 
Berglund-H\"{u}bsch-Krawitz duality}

\author{Charles F. Doran}
\author{and Richard S. Garavuso}

\affiliation{
Department of Mathematical and Statistical Sciences, University of Alberta  
\\ 
632 Central Academic Building; Edmonton, Alberta T6G 2G1; Canada
            } 

\emailAdd{doran@math.ualberta.ca}
\emailAdd{garavuso@ualberta.ca}

\abstract{
This paper discusses the overlap of the Hori-Vafa formulation of mirror symmetry with some other constructions.
We focus on compact Calabi-Yau hypersurfaces $ \mathcal{M}_G = \{ G = 0 \} $ in weighted complex projective spaces.
The Hori-Vafa formalism relates a family
$
\left\{ 
        \left.
        \mathcal{M}_G \in \mathbf{WCP}^{m-1}_{Q_1,\ldots,Q_m}[s] \,
        \right| \,
                \sum_{i=1}^m Q_i = s          
\right\}
$
of such hypersurfaces to a single Landau-Ginzburg mirror theory.
A technique suggested by Hori and Vafa allows the Picard-Fuchs equations satisfied by the corresponding mirror periods to be determined.
Some examples in which the variety $ \mathcal{M}_G $ is crepantly resolved are considered.
The resulting Picard-Fuchs equations agree with those found elsewhere working in the Batyrev-Borisov framework.
When $ G $ is an invertible nondegenerate quasihomogeneous polynomial, the 
Chiodo-Ruan geometrical interpretation of Berglund-H\"{u}bsch-Krawitz duality can be used to associate a particular complex structure for $ \mathcal{M}_G $ with a particular K\"{a}hler structure for the mirror $ \widetilde{\mathcal{M}}_G $.
We make this association for such $ G $ when 
the ambient space of $ \mathcal{M}_G $ is 
$ \mathbf{CP}^2 $, $ \mathbf{CP}^3 $, and $ \mathbf{CP}^4 $.
Finally, we probe some of the resulting mirror K\"{a}hler structures by determining corresponding Picard-Fuchs equations.
         }


\arxivnumber{}

\maketitle
\flushbottom

\section{Introduction}

A K\"{a}hler manifold with nonnegative first Chern class can be described as the resolved 
target space of a nonlinear sigma model phase of a (2,2) supersymmetric gauged linear sigma model in $ 1 + 1 $ dimensions \cite{Witten:Phases,MorrisonPlesser:Summing}.
Hori and Vafa \cite{HoriVafa:Mirror} showed that the aforementioned gauged linear sigma model is mirror to a Landau-Ginzburg theory.
This paper discusses the overlap of the Hori-Vafa formulation of mirror symmetry with some other constructions.
We focus on comapct Calabi-Yau hypersurfaces $ \mathcal{M}_G = \{ G = 0 \} $ in weighted complex projective spaces.
The Hori-Vafa formalism relates a family
\begin{equation}
\label{CY-family-WCP^m-1[s]}
\left\{ 
        \mathcal{M}_G \in \mathbf{WCP}^{m-1}_{Q_1,\ldots,Q_m}[s] \, 
        \left| \,
               \sum_{i=1}^m Q_i = s
        \right.           
\right\}
\end{equation}
of such hypersurfaces to a single Landau-Ginzburg mirror theory.
We will use a technique suggested by Hori and Vafa to obtain the Picard-Fuchs equations satisfied by the mirror periods.
Some examples in which the variety $ \mathcal{M}_G $ is crepantly resolved to yield a Calabi-Yau manifold $ M_G $ will be considered.
The resulting Picard-Fuchs equations will be compared with the results found in
\cite{HosonoKlemmTheisenYau:Mirror,LianYau:Mirror} working in the Batyrev-Borisov 
\cite{Batyrev:Dual,Borisov:Towards} framework.

The Greene-Plesser \cite{GreenePleser:Duality,GreenePlesser:Mirror} construction and its Berglund-H\"{u}bsch \cite{BerglundHuebsch:AGeneralized} generalization provides a mirror partner for each member of a special subfamily of 
(\ref{CY-family-WCP^m-1[s]}).
For the Berglund-H\"{u}bsch case, this subfamily is defined by the additional requirement that $ G $ be an invertible nondegenerate polynomial potential.
Using the ideas of Berglund and H\"{u}bsch, Krawitz \cite{Krawitz:FJRW} formulated a mirror pair construction which associates a Landau-Ginzburg orbifold
$ \mathcal{W} / \mathcal{G} $, where $ \mathcal{W} $ is an invertible nondegenerate quasihomogeneous polynomial potential and $ \mathcal{G} $ is an admissible group, with a dual Landau-Ginzburg orbifold $ \mathcal{W}^T / \mathcal{G}^T $.
This association is referred to as \emph{Berglund-H\"{u}bsch-Krawitz duality}.
Chiodo and Ruan \cite{ChiodoRuan:LG/CY} established a Landau-Ginzburg/Calabi-Yau correspondence which allowed them to obtain a geometrical interpretation of Berglund-H\"{u}bsch-Krawitz duality when 
$ \mathcal{M}_{\mathcal{W}} = \{ \mathcal{W} = 0 \} $ is a Calabi-Yau variety and
$ 
\langle J_{\mathcal{W}} \rangle 
  \subset \mathcal{G} 
  \subset \mathrm{SL}_{\mathcal{W}} 
$.  
Specifically, they established that, for such $ \mathcal{M}_{\mathcal{W}} $ and
$ \mathcal{G} $, the Calabi-Yau orbifolds
$ 
\mathcal{M}_{\mathcal{W}} /
\left( 
       \mathcal{G} / 
       \langle J_{\mathcal{W}} \rangle  
\right) 
$  
and 
$ 
\mathcal{M}_{\mathcal{W}^T} /
\left( 
       \mathcal{G}^T / 
       \langle J_{\mathcal{W}^T} \rangle  
\right)
$
form a mirror pair.
Note that when $ \mathcal{G} = \langle J_{\mathcal{W}} \rangle $, this mirror pair becomes 
$ 
\left( 
       \mathcal{M}_{\mathcal{W}} , 
       \widetilde{\mathcal{M}}_{\mathcal{W}}
\right)
$, where 
\begin{equation*}
\widetilde{\mathcal{M}}_{\mathcal{W}} 
   = \frac{ \left\{ \mathcal{\mathcal{W}}^T = 0 \right\} }
          { \mathrm{SL}_{\mathcal{W}^T} / \langle J_{\mathcal{W}^T} \rangle } \, .
\end{equation*}
We will discuss $ \mathcal{M}_{\mathcal{W}} $ and 
$ \widetilde{\mathcal{M}}_{\mathcal{W}} $ in the Hori-Vafa context and specifically consider the cases $ \mathcal{M}_{\mathcal{W}} \in \mathbf{CP}^2[3] $, 
$ \mathcal{M}_{\mathcal{W}} \in \mathbf{CP}^3[4] $, and
$ \mathcal{M}_{\mathcal{W}} \in \mathbf{CP}^4[5] $.

This paper is organized as follows:  
In Sections \ref{Compact} and \ref{Related}, we discuss the Hori-Vafa formalism pertaining to a \emph{compact} hypersurface in a toric variety and to a closely related \emph{noncompact} toric variety, respectively.
At the end of this discussion, we write down a relation between the corresponding Landau-Ginzburg mirror periods.
In Section \ref{CYPF}, we use a technique suggested by Hori and Vafa to determine the Picard-Fuchs equations satisfied by these mirror periods when a Calabi-Yau condition is satisfied.
We then specialize the discussion to consider some examples in which varieties of the form given in (\ref{CY-family-WCP^m-1[s]}) are crepantly resolved. 
In Section \ref{Geometrical}, we discuss the Chiodo-Ruan geometrical interpretation of Berglund-H\"{u}bsch-Krawitz duality in the Hori-Vafa context.
We conclude with a discussion of our results in Section \ref{Discussion}.

\section{\label{Compact}Compact hypersurface in a toric variety}
 
Let $ \mathcal{X} $ be a toric variety of complex dimension $ m - k $ defined by the charge matrix $ \left( Q_{iA} \right) $, where $ i = 1,\ldots,m $ and $ A = 1,\ldots,k $.
Consider a compact hypersurface  
\begin{equation}
\mathcal{M}_G = \left\{ G = 0 \right\} \subset \mathcal{X} \, ,
\end{equation}
where $ G = G \left( \phi_1,\ldots,\phi_m \right) $ is a quasihomogeneous polynomial of multidegree $ (s_A) $.
A K\"{a}hler manifold $ M_G $ may be obtained as a crepant resolution of $ \mathcal{M}_G $ if a crepant resolution exists.
Such a manifold would have nonnegative first Chern class when
\begin{equation}
\label{nonnegative-c_1(M_G)}
\sum_{i=1}^m Q_{iA} \geq s_A \, , \qquad A = 1,\ldots,k \, .
\end{equation}
We will assume that a crepant resolution exists and that  (\ref{nonnegative-c_1(M_G)}) is satisfied.

We can describe $ \mathcal{M}_G $ as the target space of a nonlinear sigma model phase of a 
$ (2,2) $ supersymmetric $ U(1)^k $ gauged linear sigma model with classical Lagrangian
\begin{align}
L_{\mathcal{M}_G}
\label{Lagrangian-M_G}
  &= \int d^4 \theta \,
     \left(
            \sum_{i=1}^m \overline{\Phi}_i e^{2 \sum_{A=1}^k Q_{iA} V_A} \Phi_i
          + \overline{P} e^{-2 \sum_{A=1}^k s_A V_A} P
          + \sum_{A,B=1}^k \frac{1}{2 e^2_{AB}} \overline{\Sigma}_A \Sigma_B   
     \right)
\nonumber
\\
  &\phantom{=}
   - \frac{1}{2}
     \left(
            \int d^2 \tilde{\theta} \; \sum_{A=1}^k t_A \Sigma_A + c.c.  
     \right)
   + \left(
            \int d^2 \theta \; 
            P \cdot G \left( \Phi_1,\ldots,\Phi_m \right) + c.c.
     \right).          
\end{align}
Here, $ \Sigma_A $ is the twisted chiral field strength of the U(1) vector superfield $ V_A $,
$ \Phi_i $ and $ P $ are chiral superfields of respective charges $ Q_{iA} $ and $ -s_A $ under the $ A $-th $ U(1) $, $ G = G \left( \Phi_1,\ldots,\Phi_m \right) $ is a quasihomogeneous polynomial of charge $ s_A $ under the $ A $-th $ U(1) $, $ e_{AB} $ is the gauge coupling, and $ t_A = r_A - i \vartheta_A $ is a complexified Fayet-Iliopoulos parameter.
The nonlinear sigma model phase is realized in the low energy limit with
\begin{equation*}
r_A \gg 0 \, , \qquad \sigma_A = 0 \, , \qquad p = 0 \, ,
\end{equation*}
and target space
\begin{equation}
\mathcal{M}_G 
   = \left\{ G = 0 \right\}
   \subset 
   \frac{ \left\{ 
                  \left( \phi_1,\cdots,\phi_m \right) \,
                  \left| \,
                         \sum_{i=1}^m Q_{iA} | \phi_i |^2 = r_A \, , \
                         A=1,\ldots,k
                  \right.
          \right\}                         
        }{U(1)^k} \, ,
\end{equation}
where $ \sigma_A $, $ \phi_i $ and $ p $ are the lowest components of the $ \theta $-expansions of $ \Sigma_A $, $ \Phi_i $, and $ P $, respectively.
If the Calabi-Yau condition
\begin{equation}
\label{c_1(M_G)=0}
\sum_{i=1}^m Q_{iA} = s_A \, , \qquad A = 1,\ldots,k 
\end{equation} 
is satisfied, then each Fayet-Iliopoulos parameter $ r_A $ does not renormalize.
Following \cite{HoriVafa:Mirror}, we obtain the Landau-Ginzburg mirror period 
\begin{align}
\Pi_{\widetilde{\mathcal{M}}_G}
\label{mirror-period-M_G}
  &= \int \left( \prod_{i=1}^m dY_i \right) dY_P \, e^{-Y_P}
     \left[
            \prod_{A=1}^k 
            \delta \left(
                          \sum_{i=1}^m Q_{iA} Y_i - s_A Y_P - t_A
                   \right)    
     \right]
\nonumber
\\
  &\phantom{= \int} \times     
     \exp{ \left(
                - \sum_{i=1}^m e^{-Y_i} - e^{-Y_P} 
           \right)   
         } \, ,  
\end{align}
where $ Y_i $ and $ Y_P $ are the dual variables of $ \Phi_i $ and $ \Phi_P $, respectively.

\section{\label{Related}Related noncompact toric variety}
 
The compact variety $ \mathcal{M}_G = \{ G = 0 \} \subset \mathcal{X} $ discussed in the previous section is closely related to the noncompact toric variety 
$ 
\mathcal{N} 
   = \mathrm{Tot} 
     \left( 
            \oplus_{A=1}^k \mathcal{O}(-s_A) \rightarrow \mathcal{X} 
     \right) 
$ 
defined by the charge matrix $ (Q_{iA}|-s_A) $.
We can describe $ \mathcal{N} $ as the target space of a nonlinear sigma model phase of a 
$ (2,2) $ supersymmetric $ U(1)^k $ gauged linear sigma model with classical Lagrangian
given by (\ref{Lagrangian-M_G}) without the superpotential term, i.e.
\begin{align}
L_{\mathcal{N}}
  &= \int d^4 \theta \,
     \left(
            \sum_{i=1}^m \overline{\Phi}_i e^{2 \sum_{A=1}^k Q_{iA} V_A} \Phi_i
          + \overline{P} e^{-2 \sum_{A=1}^k s_A V_A} P
          + \sum_{A,B=1}^k \frac{1}{2 e^2_{AB}} \overline{\Sigma}_A \Sigma_B   
     \right)
\nonumber
\\
  &\phantom{=}
   - \frac{1}{2}
     \left(
            \int d^2 \tilde{\theta} \; \sum_{A=1}^k t_A \Sigma_A + c.c.  
     \right).          
\end{align}
The nonlinear sigma model phase is realized in the low energy limit with
\begin{equation*}
r_A \gg 0 \, , \qquad \sigma_A = 0 \, ,
\end{equation*}
and target space
\begin{equation}
\mathcal{N} 
   = \frac{ 
            \left\{ 
                    \left( \phi_1,\cdots,\phi_m,p \right) \,
                           \left| \,
                                  \sum_{i=1}^m Q_{iA} | \phi_i |^2  - s_A |p|^2 = r_A \, , \
                                  A=1,\ldots,k
                           \right.
            \right\}                         
          }{U(1)^k} \, .
\end{equation}
Following \cite{HoriVafa:Mirror}, we obtain the Landau-Ginzburg mirror period
\begin{align}
\Pi_{\widetilde{\mathcal{N}}}
\label{mirror-period-N}
  &= \int \left( \prod_{i=1}^m dY_i \right) dY_P
     \left[
            \prod_{A=1}^k 
            \delta \left(
                          \sum_{i=1}^m Q_{iA} Y_i - s_A Y_P - t_A
                   \right)    
     \right]  
     \exp{ \left(
                - \sum_{i=1}^m e^{-Y_i} - e^{-Y_P} 
           \right)   
         } \, .
\end{align} 
The periods (\ref{mirror-period-M_G}) and (\ref{mirror-period-N}) are related by
\begin{equation}
\label{mirror-period-relation-M_G-N}
\Pi_{\widetilde{\mathcal{M}}_G}
   = - \left(
              \sum_{A=1}^k s_A \frac{\partial}{\partial t_A} 
       \right) 
       \Pi_{\widetilde{\mathcal{N}}} \, .
\end{equation}
\section{\label{CYPF}Calabi-Yau Picard-Fuchs equations}

In this section, we will use a technique suggested in \cite{HoriVafa:Mirror} to determine the Picard-Fuchs equations satisfied by $ \Pi_{\widetilde{\mathcal{M}}_G} $ when the Calabi-Yau condition (\ref{c_1(M_G)=0}) holds.
This technique involves first determining the Picard-Fuchs equations satisfied by 
$ \Pi_{\widetilde{\mathcal{N}}} $ when the Calabi-Yau condition holds and then using the relation (\ref{mirror-period-relation-M_G-N}).

We begin by considering
\begin{align}
\label{mirror-period-N(mu,t)}
\Pi_{\widetilde{\mathcal{N}}} (\textrm{\boldmath $ \mu $}, \mathbf{t})
  &= \int \left( \prod_{i=1}^m dY_i \right) dY_P
     \left[
            \prod_{A=1}^k 
            \delta \left(
                          \sum_{i=1}^m Q_{iA} Y_i - s_A Y_P - t_A
                   \right)    
     \right]
\nonumber
\\
  &\phantom{= \int} \times       
     \exp{ \left(
                - \sum_{i=1}^m \mu_i e^{-Y_i} - \mu_P e^{-Y_P} 
           \right)   
         } \, ,
\end{align} 
where
\begin{equation}
\textrm{\boldmath $ \mu $} = \left( \mu_1,\ldots,\mu_m,\mu_P \right) \, ,
\qquad
\mathbf{t} = \left( t_1,\ldots,t_k \right).   
\end{equation}
When the Calabi-Yau condition (\ref{c_1(M_G)=0}) holds, 
$ \Pi_{\widetilde{\mathcal{N}}} (\textrm{\boldmath $ \mu $}, \mathbf{t}) $ satisfies
\begin{equation}
\label{GKZ-mirror-period-N(mu,t)}
\prod_{Q_{iA} > 0} \left( \frac{\partial}{\partial \mu_i} \right)^{Q_{iA}} 
\Pi_{\widetilde{\mathcal{N}}} (\textrm{\boldmath $ \mu $}, \mathbf{t}) 
   = e^{-t_A} \left( \frac{\partial}{\partial \mu_P} \right)^{s_A}
     \prod_{Q_{iA} < 0} \left( \frac{\partial}{\partial \mu_i} \right)^{-Q_{iA}}
     \Pi_{\widetilde{\mathcal{N}}} (\textrm{\boldmath $ \mu $}, \mathbf{t}) \, ,
\end{equation}
for $ A = 1,\ldots,k $.
Under the shifts
\begin{equation}
Y_i \rightarrow Y_i + \ln{\mu_i} \, ,
\quad
i = 1,\ldots,m \, ;
\qquad
Y_P \rightarrow Y_P + \ln{\mu_P} \, ,
\end{equation}
we can eliminate the {\boldmath $ \mu $} dependence in (\ref{mirror-period-N(mu,t)}), except for a shift in each delta function constraint.
That is,
\begin{align}
\label{mirror-period-N(mu,t)=N(t)}
\Pi_{\widetilde{\mathcal{N}}} (\textrm{\boldmath $ \mu $}, \mathbf{t})
  &= \Pi_{\widetilde{\mathcal{N}}} (\mathbf{T})    
\nonumber
\\
  &= \int \left( \prod_{i=1}^m dY_i \right) dY_P
     \left[
            \prod_{A=1}^k 
            \delta \left(
                          \sum_{i=1}^m Q_{iA} Y_i - s_A Y_P - T_A
                   \right)    
     \right]
\nonumber
\\
  &\phantom{= \int} \times        
     \exp{ \left(
                - \sum_{i=1}^m e^{-Y_i} - e^{-Y_P} 
           \right)   
         } \, ,
\end{align}
where
\begin{equation}
\label{T_A}
\mathbf{T} = (T_1,...,T_k) \, ;
\qquad
T_A = t_A - \sum_{j=1}^m Q_{jA} \ln{\mu_j} + s_A \ln{\mu_P} \, ,
\quad
A = 1,\ldots,k \, .
\end{equation} 
Using (\ref{mirror-period-N(mu,t)=N(t)}) in (\ref{GKZ-mirror-period-N(mu,t)}) gives
\begin{equation}
\label{GKZ-mirror-period-N(T)}
\prod_{Q_{iA} > 0} \left( \frac{\partial}{\partial \mu_i} \right)^{Q_{iA}} 
\Pi_{\widetilde{\mathcal{N}}} (\mathbf{T})
   = e^{-t_A} \left( \frac{\partial}{\partial \mu_P} \right)^{s_A}
     \prod_{Q_{iA} < 0} \left( \frac{\partial}{\partial \mu_i} \right)^{-Q_{iA}}
     \Pi_{\widetilde{\mathcal{N}}} (\mathbf{T}) \, ,  
\end{equation}
for $ A = 1,\ldots,k $.
From the chain rule, we have
\begin{equation}
\begin{aligned}
\label{chain-rule}
\frac{\partial}{\partial \mu_i} \Pi_{\widetilde{\mathcal{N}}} (\mathbf{T})
  &= \left(
            \sum_{A=1}^k \frac{Q_{iA}}{\mu_i} \Theta_A
     \right)
     \Pi_{\widetilde{\mathcal{N}}} (\mathbf{T}) \, ,
\qquad
i=1,\ldots,m \, ;     
\\
\frac{\partial}{\partial \mu_P} \Pi_{\widetilde{\mathcal{N}}} (\mathbf{T})
  &= - \left(
              \sum_{A=1}^k \frac{s_A}{\mu_P} \Theta_A
       \right)
       \Pi_{\widetilde{\mathcal{N}}} (\mathbf{T}) \, ,
\end{aligned}
\end{equation}
where 
\begin{equation}
\Theta_A \equiv - \frac{\partial}{\partial T_A} \, ,
\qquad
A = 1,\ldots,k \, .
\end{equation}
Furthermore, from (\ref{T_A}), we obtain
\begin{equation}
\label{mu-T-t-relation}
\frac{1}{ \mu_1^{Q_{1A}} \cdots \mu_m^{Q_{mA}} }
   = \frac{e^{T_A} e^{-t_A}}{\mu_P^{s_A}} \, ,
\qquad   
A = 1,\ldots,k \, .
\end{equation}
Equations (\ref{chain-rule}) and (\ref{mu-T-t-relation}) can be used to eliminate 
$ \mu_1,\ldots,\mu_m,\mu_P $ and $ t_A $ from (\ref{GKZ-mirror-period-N(T)}). 
Doing this for $ A = 1,\ldots,k $ and then making the replacements
\begin{equation}
\label{replacements}
T_A \rightarrow t_A \, , 
\qquad
\Theta_A \rightarrow \theta_A \equiv - \frac{\partial}{\partial t_A} \, ;
\qquad
A = 1,\ldots,k 
\end{equation}
yields the Picard-Fuchs equations satisfied by $ \Pi_{\widetilde{\mathcal{N}}} $.
Applying (\ref{mirror-period-relation-M_G-N}) then allows the Picard-Fuchs equations satisfied by $ \Pi_{\widetilde{\mathcal{M}}_G} $ to be determined.
This procedure can also be applied to the corresponding crepantly resolved manifolds
$ \widetilde{M}_G $ and $ \widetilde{N} $.
We will now illustrate this with some examples.

\begin{example}

\label{PF-Example1}
Let $ \mathcal{M}_G \in \mathbf{WCP}^4_{1,1,2,2,2}[8] $.
This singular variety is described by the charge matrix
$ (Q_1, Q_2, Q_3, Q_4, Q_5 \, | -s) = (1,1,2,2,2 \, | -8) $.
The singularities can be crepantly resolved to yield $ M_G $, which is described by the charge matrix 
\begin{equation*}
\left( Q_{iA} | -s_A \right) 
   = \left(
            \begin{array}{cccccc|c}
               Q_{11} & Q_{21} & Q_{31} & Q_{41} & Q_{51} & Q_{61} & -s_1 \\
               Q_{12} & Q_{22} & Q_{32} & Q_{42} & Q_{52} & Q_{62} & -s_2 
            \end{array}
     \right)       
   = \left(
            \begin{array}{cccccc|c}
               0 & 0 & 1 & 1 & 1 & 1 & -4 \\
               1 & 1 & 0 & 0 & 0 & -2 & 0 
            \end{array}
     \right).  
\end{equation*}
Note that
\begin{align*} 
\Pi_{\widetilde{M}_G}
  &= - 4 \frac{\partial}{\partial t_1}  \Pi_{\widetilde{N}}
\nonumber
\\
  &= \int \left( \prod_{i=1}^6 dY_i \right) dY_P \, e^{-Y_P}  
     \delta \left(
                   Y_3 + Y_4 + Y_5 + Y_6 - 4 Y_P - t_1
           \right)    
\nonumber
\\
  &\phantom{= \int} \times
     \delta \left(
                   Y_1 + Y_2 - 2 Y_6 - t_2
           \right)      
     \exp{ \left(
                - \sum_{i=1}^6 e^{-Y_i} - e^{-Y_P} 
           \right)   
         } \, ,    
\end{align*}
where
\begin{align*}
\Pi_{\widetilde{N}}
  &= \int \left( \prod_{i=1}^6 dY_i \right) dY_P \,  
     \delta \left(
                   Y_3 + Y_4 + Y_5 + Y_6 - 4 Y_P - t_1
           \right)    
\nonumber
\\
  &\phantom{= \int} \times
     \delta \left(
                   Y_1 + Y_2 - 2 Y_6 - t_2
           \right)      
     \exp{ \left(
                - \sum_{i=1}^6 e^{-Y_i} - e^{-Y_P} 
           \right)   
         } \, . 
\end{align*}
Equation (\ref{GKZ-mirror-period-N(T)}) then yields
\begin{align*}
\frac{\partial}{\partial \mu_3}
\frac{\partial}{\partial \mu_4}
\frac{\partial}{\partial \mu_5}
\frac{\partial}{\partial \mu_6}
\Pi_{\widetilde{N}} (\mathbf{T})
  &= e^{-t_1} \frac{\partial^4}{\partial \mu_P^4}
     \Pi_{\widetilde{N}} (\mathbf{T}) \, ,
\\[1ex]
\frac{\partial}{\partial \mu_1}
\frac{\partial}{\partial \mu_2}
\Pi_{\widetilde{N}} (\mathbf{T})
  &= e^{-t_2} \frac{\partial^2}{\partial \mu_6^2}
     \Pi_{\widetilde{N}} (\mathbf{T}) \, ,    
\end{align*}
for $ A = 1 $ and $ A = 2 $, respectively.
Using (\ref{chain-rule}), we obtain
\begin{align*}
\frac{1}{\mu_3 \mu_4 \mu_5 \mu_6}
\Theta_1^3 (\Theta_1 - 2 \Theta_2) \, 
\Pi_{\widetilde{N}} (\mathbf{T})
  &= \frac{e^{-t_1}}{\mu_P^4} 
     (4 \Theta_1 + 3) (4 \Theta_1 + 2) (4 \Theta_1 + 1) (4 \Theta_1) \, 
     \Pi_{\widetilde{N}} (\mathbf{T}) \, ,
\\[1ex]
\frac{1}{\mu_1 \mu_2}
\Theta_2^2 \,
\Pi_{\widetilde{N}} (\mathbf{T})
  &= \frac{e^{-t_2}}{\mu_6^2}  
     (2 \Theta_2 - \Theta_1 + 1) (2 \Theta_2 - \Theta_1) \,
     \Pi_{\widetilde{N}} (\mathbf{T}) \, .  
\end{align*}
Applying (\ref{mu-T-t-relation}) and rearranging gives
\begin{align*}
0 &= \left[
            \frac{1}{4} \Theta_1^2 (\Theta_1 - 2 \Theta_2)
          - e^{-T_1} (4 \Theta_1 + 3) (4 \Theta_1 + 2) (4 \Theta_1 + 1)
     \right] 4 \Theta_1 \,
     \Pi_{\widetilde{N}} (\mathbf{T}) \, ,
\\[1ex]
0 &= \left[
            \Theta_2^2 
          - e^{-T_2} (2 \Theta_2 - \Theta_1 + 1) (2 \Theta_2 - \Theta_1)
     \right] 
     \Pi_{\widetilde{N}} (\mathbf{T}) \, .    
\end{align*}          
Making the replacements (\ref{replacements}) yields
\begin{align*}
0 &= \left[
            \frac{1}{4} \theta_1^2 \left( \theta_1 - 2 \theta_2 \right)
          - e^{-t_1} (4 \theta_1 + 3) (4 \theta_1 + 2) (4 \theta_1 + 1)
     \right] 4 \theta_1 \,
     \Pi_{\widetilde{N}} \, ,
\\[1ex]
0 &= \left[
            \theta_2^2 
          - e^{-t_2} (2 \theta_2 - \theta_1 + 1) (2 \theta_2 - \theta_1)
     \right] 
     \Pi_{\widetilde{N}} \, ,    
\end{align*}
which are the Picard-Fuchs equations satisfied by $ \Pi_{\widetilde{N}} $.
Finally, using the relation
\begin{equation*}
\Pi_{\widetilde{M}_G}
   = 4 \theta_1 \, \Pi_{\widetilde{N}} \, ,
\end{equation*}
we obtain the Picard-Fuchs equations satisfied by $ \Pi_{\widetilde{M}_G} $, i.e.
\begin{equation}
\label{Picard-Fuchs-WCP^4_(1,1,2,2,2)[8]}
\begin{aligned}
0 &= \left[
            \theta_1^2 \left( \theta_1 - 2 \theta_2 \right)
          - 4 e^{-t_1} (4 \theta_1 + 3) (4 \theta_1 + 2) (4 \theta_1 + 1)
     \right]
     \Pi_{\widetilde{M}_G} \, ,
\\[1ex]
0 &= \left[
            \theta_2^2 
          - e^{-t_2} (2 \theta_2 - \theta_1 + 1) (2 \theta_2 - \theta_1)
     \right] 
     \Pi_{\widetilde{M}_G} \, .   
\end{aligned}
\end{equation}
Note that our result (\ref{Picard-Fuchs-WCP^4_(1,1,2,2,2)[8]}) agrees with the result in
\cite{HosonoKlemmTheisenYau:Mirror} obtained via a different method.

\end{example}

\begin{example}

\label{PF-Example2}
Let $ \mathcal{M}_G \in \mathbf{WCP}^4_{6,2,2,1,1}[12] $.
This singular variety is described by the charge matrix
$ (Q_1, Q_2, Q_3, Q_4, Q_5 \, | -s) = (6,2,2,1,1 \, | -12) $.
The singularities can be crepantly resolved to yield $ M_G $, which is described by the charge matrix 
\begin{equation*}
\left( Q_{iA} | -s_A \right) 
   = \left(
            \begin{array}{cccccc|c}
               Q_{11} & Q_{21} & Q_{31} & Q_{41} & Q_{51} & Q_{61} & -s_1 \\
               Q_{12} & Q_{22} & Q_{32} & Q_{42} & Q_{52} & Q_{62} & -s_2 
            \end{array}
     \right)       
   = \left(
            \begin{array}{cccccc|c}
               3 & 1 & 1 & 0 & 0 & 1 & -6 \\
               0 & 0 & 0 & 1 & 1 & -2 & 0 
            \end{array}
     \right).  
\end{equation*}
Note that
\begin{align*} 
\Pi_{\widetilde{M}_G}
  &= - 6 \frac{\partial}{\partial t_1}  \Pi_{\widetilde{N}}
\nonumber
\\
  &= \int \left( \prod_{i=1}^6 dY_i \right) dY_P \, e^{-Y_P}  
     \delta \left(
                   3 Y_1 + Y_2 + Y_3 + Y_6 - 6 Y_P - t_1
           \right)    
\nonumber
\\
  &\phantom{= \int} \times
     \delta \left(
                   Y_4 + Y_5 - 2 Y_6 - t_2
           \right)      
     \exp{ \left(
                - \sum_{i=1}^6 e^{-Y_i} - e^{-Y_P} 
           \right)   
         } \, ,    
\end{align*}
where
\begin{align*}
\Pi_{\widetilde{N}}
  &= \int \left( \prod_{i=1}^6 dY_i \right) dY_P \,  
     \delta \left(
                   3 Y_1 + Y_2 + Y_3 + Y_6 - 6 Y_P - t_1
           \right)    
\nonumber
\\
  &\phantom{= \int} \times
     \delta \left(
                   Y_4 + Y_5 - 2 Y_6 - t_2
           \right)      
     \exp{ \left(
                - \sum_{i=1}^6 e^{-Y_i} - e^{-Y_P} 
           \right)   
         } \, . 
\end{align*}
Equation (\ref{GKZ-mirror-period-N(T)}) then yields
\begin{align*}
\frac{\partial^3}{\partial \mu_1^3}
\frac{\partial}{\partial \mu_2}
\frac{\partial}{\partial \mu_3}
\frac{\partial}{\partial \mu_6}
\Pi_{\widetilde{N}} (\mathbf{T})
  &= e^{-t_1} \frac{\partial^6}{\partial \mu_P^6}
     \Pi_{\widetilde{N}} (\mathbf{T}) \, ,
\\[1ex]
\frac{\partial}{\partial \mu_4}
\frac{\partial}{\partial \mu_5}
\Pi_{\widetilde{N}} (\mathbf{T})
  &= e^{-t_2} \frac{\partial^2}{\partial \mu_6^2}
     \Pi_{\widetilde{N}} (\mathbf{T}) \, ,    
\end{align*}
for $ A = 1 $ and $ A = 2 $, respectively.
Using (\ref{chain-rule}), we obtain
\begin{multline*}
\frac{1}{\mu_1^3 \mu_2 \mu_3 \mu_6}
(3 \Theta_1 - 2) (3 \Theta_1 - 1) (3 \Theta_1)
\Theta_1^2 (\Theta_1 - 2 \Theta_2) \,
\Pi_{\widetilde{N}} (\mathbf{T})
\nonumber
\\
  = 8 \frac{e^{-t_1}}{\mu_P^6} 
    (6 \Theta_1 + 5) (3 \Theta_1 + 2) (6 \Theta_1 + 3) (3 \Theta_1 + 1) (6 \Theta_1 + 1)
    (3 \Theta_1) \,
    \Pi_{\widetilde{N}} (\mathbf{T}) \, ,
\end{multline*}
\begin{equation*}
\frac{1}{\mu_4 \mu_5}
\Theta_2^2 \,
\Pi_{\widetilde{N}} (\mathbf{T})
   = \frac{e^{-t_2}}{\mu_6^2}  
     (2 \Theta_2 - \Theta_1 + 1) (2 \Theta_2 - \Theta_1)
     \Pi_{\widetilde{N}} (\mathbf{T}) \, .       
\end{equation*}
Applying (\ref{mu-T-t-relation}) and rearranging gives
\begin{align*}
0 &= (3 \Theta_1 - 2) (3 \Theta_1 - 1)
     \left[
            \Theta_1^2 (\Theta_1 - 2 \Theta_2)
     \right.       
\\
  &\phantom{=}
     \left.               
          - 8 e^{-T_1} (6 \Theta_1 + 5) (6 \Theta_1 + 3) (6 \Theta_1 + 1)
     \right] 3\Theta_1 \,
     \Pi_{\widetilde{N}} (\mathbf{T}) \, ,
\\[1ex]
0 &= \left[
            \Theta_2^2 
          - e^{-T_2} (2 \Theta_2 - \Theta_1 + 1) (2 \Theta_2 - \Theta_1)
     \right] 
     \Pi_{\widetilde{N}} (\mathbf{T}) \, ,   
\end{align*}
where we have used the identity
\begin{equation*}
e^{-T_1} (\Theta_1 + 1) \, \Pi_{\widetilde{N}} (\mathbf{T})
   = \Theta_1 e^{-T_1} \, \Pi_{\widetilde{N}} (\mathbf{T}) \, .
\end{equation*}          
Making the replacements (\ref{replacements}) yields
\begin{align*}
0 &= (3 \theta_1 - 2) (3 \theta_1 - 1)
     \left[
            \theta_1^2 (\theta_1 - 2 \theta_2)            
          - 8 e^{-t_1} (6 \theta_1 + 5) (6 \theta_1 + 3) (6 \theta_1 + 1)
     \right] 3 \theta_1 \,
     \Pi_{\widetilde{N}} \, ,
\\[1ex]
0 &= \left[
            \theta_2^2 
          - e^{-t_2} (2 \theta_2 - \theta_1 + 1) (2 \theta_2 - \theta_1)
     \right] 
     \Pi_{\widetilde{N}} \, ,   
\end{align*}
which are the Picard-Fuchs equations satisfied by $ \Pi_{\widetilde{N}} $.
Finally, using the relation
\begin{equation*}
\Pi_{\widetilde{M}_G}
   = 6 \theta_1 \, \Pi_{\widetilde{N}}
\end{equation*}
and removing the factor $ (3 \theta_1 - 2) (3 \theta_1 - 1) $,
we obtain the Picard-Fuchs equations satisfied by $ \Pi_{\widetilde{M}_G} $, i.e.
\begin{equation}
\label{Picard-Fuchs-WCP^4_(6,2,2,1,1)[12]}
\begin{aligned}
0 &= \left[
            \theta_1^2 (\theta_1 - 2 \theta_2)            
          - 8 e^{-t_1} (6 \theta_1 + 5) (6 \theta_1 + 3) (6 \theta_1 + 1)
     \right]
     \Pi_{\widetilde{M}_G} \, ,
\\[1ex]
0 &= \left[
            \theta_2^2 
          - e^{-t_2} (2 \theta_2 - \theta_1 + 1) (2 \theta_2 - \theta_1)
     \right] 
     \Pi_{\widetilde{M}_G} \, .   
\end{aligned}
\end{equation}
Note that our result (\ref{Picard-Fuchs-WCP^4_(6,2,2,1,1)[12]}) agrees with the result in
\cite{HosonoKlemmTheisenYau:Mirror} obtained via a different method.

\end{example}

\begin{example}

\label{PF-Example3}
Let $ \mathcal{M}_G \in \mathbf{WCP}^3_{1,1,4,6}[12] $.
This singular variety is described by the charge matrix
$ (Q_1, Q_2, Q_3, Q_4 \, | -s) = (1,1,4,6 \, | -12) $.
The singularities can be crepantly resolved to yield $ M_G $, which is described by the charge matrix
\begin{equation*}
\left( Q_{iA} | -s_A \right) 
   = \left(
            \begin{array}{ccccc|c}
               Q_{11} & Q_{21} & Q_{31} & Q_{41} & Q_{51} & -s_1 \\
               Q_{12} & Q_{22} & Q_{32} & Q_{42} & Q_{52} & -s_2 
            \end{array}
     \right)       
   = \left(
            \begin{array}{ccccc|c}
               0 & 0 & 2 & 3 & 1 & -6 \\
               1 & 1 & 0 & 0 & -2 & 0 
            \end{array}
     \right).  
\end{equation*}
Note that
\begin{align*} 
\Pi_{\widetilde{M}_G}
  &= - 6 \frac{\partial}{\partial t_1}  \Pi_{\widetilde{N}}
\nonumber
\\
  &= \int \left( \prod_{i=1}^5 dY_i \right) dY_P \, e^{-Y_P}  
     \delta \left(
                   2 Y_3 + 3 Y_4 + Y_5 - 6 Y_P - t_1
           \right)    
\nonumber
\\
  &\phantom{= \int} \times
     \delta \left(
                   Y_1 + Y_2 - 2 Y_5 - t_2
           \right)      
     \exp{ \left(
                - \sum_{i=1}^5 e^{-Y_i} - e^{-Y_P} 
           \right)   
         } \, ,    
\end{align*}
where
\begin{align*}
\Pi_{\widetilde{N}}
  &= \int \left( \prod_{i=1}^5 dY_i \right) dY_P \,  
     \delta \left(
                   2 Y_3 + 3 Y_4 + Y_5 - 6 Y_P - t_1
           \right)    
\nonumber
\\
  &\phantom{= \int} \times
     \delta \left(
                   Y_1 + Y_2 - 2 Y_5 - t_2
           \right)      
     \exp{ \left(
                - \sum_{i=1}^5 e^{-Y_i} - e^{-Y_P} 
           \right)   
         } \, . 
\end{align*}
Equation (\ref{GKZ-mirror-period-N(T)}) then yields
\begin{align*}
\frac{\partial^2}{\partial \mu_3^2}
\frac{\partial^3}{\partial \mu_4^3}
\frac{\partial}{\partial \mu_5}
\Pi_{\widetilde{N}} (\mathbf{T})
  &= e^{-t_1} \frac{\partial^6}{\partial \mu_P^6}
     \Pi_{\widetilde{N}} (\mathbf{T}) \, ,
\\[1ex]
\frac{\partial}{\partial \mu_1}
\frac{\partial}{\partial \mu_2}
\Pi_{\widetilde{N}} (\mathbf{T})
  &= e^{-t_2} \frac{\partial^2}{\partial \mu_5^2}
     \Pi_{\widetilde{N}} (\mathbf{T}) \, ,    
\end{align*}
for $ A = 1 $ and $ A = 2 $, respectively.
Using (\ref{chain-rule}), we obtain
\begin{multline*}
\frac{1}{\mu_3^2 \mu_4^3 \mu_5}
(2 \Theta_1 - 1) (2 \Theta_1) (3 \Theta_1 - 2) (3 \Theta_1 - 1) (3 \Theta_1)
(\Theta_1 - 2 \Theta_2) \, 
\Pi_{\widetilde{N}} (\mathbf{T})
\\
   = 8 \frac{e^{-t_1}}{\mu_P^6} 
     (6 \Theta_1 + 5) (3 \Theta_1 + 2) (6 \Theta_1 + 3) (3 \Theta_1 + 1) (6 \Theta_1 + 1) 
     (3 \Theta_1) \, 
     \Pi_{\widetilde{N}} (\mathbf{T}) \, ,
\end{multline*}
\begin{equation*}
\frac{1}{\mu_1 \mu_2}
\Theta_2^2 \,
\Pi_{\widetilde{N}} (\mathbf{T})
   = \frac{e^{-t_2}}{\mu_5^2}  
     (2 \Theta_2 - \Theta_1 + 1) (2 \Theta_2 - \Theta_1) \,
     \Pi_{\widetilde{N}} (\mathbf{T}) \, .  
\end{equation*}
Applying (\ref{mu-T-t-relation}) and rearranging gives
\begin{align*}
0 &= (2 \Theta_1 - 1) (3 \Theta_1 - 2) (3 \Theta_1 - 1)
     \left[
            \Theta_1 (\Theta_1 - 2 \Theta_2)
     \right.       
\\            
  &\phantom{=}
     \left.  
           - 12 e^{-T_1} (6 \Theta_1 + 5) (6 \Theta_1 + 1)
     \right] 6 \Theta_1 \,
     \Pi_{\widetilde{N}} (\mathbf{T}) \, ,
\\[1ex]
0 &= \left[
            \Theta_2^2 
          - e^{-T_2} (2 \Theta_2 - \Theta_1 + 1) (2 \Theta_2 - \Theta_1)
     \right] 
     \Pi_{\widetilde{N}} (\mathbf{T}) \, ,   
\end{align*}
where we have used the identity
\begin{equation*}
e^{-T_1} (\Theta_1 + 1) \, \Pi_{\widetilde{N}} (\mathbf{T})
   = \Theta_1 e^{-T_1} \, \Pi_{\widetilde{N}} (\mathbf{T}) \, .
\end{equation*}          
Making the replacements (\ref{replacements}) yields
\begin{align*}
0 &= (2 \theta_1 - 1) (3 \theta_1 - 2) (3 \theta_1 - 1)
     \left[
            \theta_1 (\theta_1 - 2 \theta_2)
          - 12 e^{-t_1} (6 \theta_1 + 5) (6 \theta_1 + 1)  
     \right] 6\theta_1 \,
     \Pi_{\widetilde{N}} \, ,
\\[1ex]
0 &= \left[
            \theta_2^2 
          - e^{-t_2} (2 \theta_2 - \theta_1 + 1) (2 \theta_2 - \theta_1)
     \right] 
     \Pi_{\widetilde{N}} \, ,    
\end{align*}
which are the Picard-Fuchs equations satisfied by $ \Pi_{\widetilde{N}} $.
Finally, using the relation
\begin{equation*}
\Pi_{\widetilde{M}_G}
   = 6 \theta_1 \, \Pi_{\widetilde{N}}
\end{equation*}
and removing the factor $ (2 \theta_1 - 1) (3 \theta_1 - 2) (3 \theta_1 - 1) $,
we obtain the Picard-Fuchs equations satisfied by $ \Pi_{\widetilde{M}_G} $, i.e.
\begin{equation}
\label{Picard-Fuchs-WCP^3_(1,1,4,6)[12]}
\begin{aligned}
0 &= \left[
            \theta_1 (\theta_1 - 2 \theta_2)
          - 12 e^{-t_1} (6 \theta_1 + 5) (6 \theta_1 + 1) 
     \right]
     \Pi_{\widetilde{M}_G} \, ,
\\[1ex]
0 &= \left[
            \theta_2^2 
          - e^{-t_2} (2 \theta_2 - \theta_1 + 1) (2 \theta_2 - \theta_1)
     \right] 
     \Pi_{\widetilde{M}_G} \, .   
\end{aligned}
\end{equation}
Note that our result (\ref{Picard-Fuchs-WCP^3_(1,1,4,6)[12]}) agrees with the result in
\cite{LianYau:Mirror} obtained via a different method.

\end{example}

Let us consider the unresolved Calabi-Yau varieties $ \mathcal{M}_G $ given at the start of Examples \ref{PF-Example1}, \ref{PF-Example2}, and \ref{PF-Example3}.
The corresponding Picard-Fuchs equations satisfied by $ \Pi_{\widetilde{\mathcal{M}}_G} $ are
\begin{align}
0 &= \left[
            (2 \theta - 1)^2 \theta^4              
          + 4 \, e^{-t} 
           (8 \theta + 7) (8 \theta + 6) (8 \theta + 5) (8 \theta + 3) (8 \theta + 2)     
           (8 \theta + 1)
     \right]
   \Pi_{\widetilde{\mathbf{WCP}}^4_{1,1,2,2,2}[8]} \, ,
\\[1ex]
0 &= \left[
            (2 \theta - 1)^2 \theta^4             
          + 16 \, e^{-t} 
           (12 \theta + 11) (12 \theta + 9) (12 \theta + 7) (12 \theta + 5) (12 \theta + 3)
           (12 \theta + 1)
     \right]
\nonumber
\\
  &\phantom{=[}      
     \Pi_{\widetilde{\mathbf{WCP}}^4_{6,2,2,1,1}[12]} \, ,
\\[1ex]
0 &= \left[
            (2 \theta - 1) \theta^3              
          + 72 \, e^{-t} 
           (12 \theta + 11) (12 \theta + 7) (12 \theta + 5) (12 \theta + 1)
     \right]
   \Pi_{\widetilde{\mathbf{WCP}}^3_{1,1,4,6}[12]} \, .                 
\end{align}

\section{\label{Geometrical}Geometrical interpretation of the mirror}

When the Calabi-Yau condition (\ref{c_1(M_G)=0}) is satisfied, we expect to find a geometrical interpretation of the Landau-Ginzburg mirror theory.  
In this section, we will discuss the realization of this expectation when $ \mathcal{M}_G $ is a Calabi-Yau hypersurface in a weighted complex projective space and $ G $ is an invertible nondegenerate quasihomogeneous polynomial.
These restrictions allow us to discuss the Chiodo-Ruan geometrical interpretation of Berglund-H\"{u}bsch-Krawitz duality in the Hori-Vafa context.

\subsection{Berglund-H\"{u}bsch-Krawitz duality}

Using the ideas of Berglund and H\"{u}bsch \cite{BerglundHuebsch:AGeneralized}, Krawitz \cite{Krawitz:FJRW} formulated a mirror pair construction which associates a Landau-Ginzburg orbifold $ \mathcal{W} / \mathcal{G} $, where $ \mathcal{W} $ is an invertible nondegenerate quasihomogeneous polynomial potential and $ \mathcal{G} $ is an admissible group, with a dual Landau-Ginzburg orbifold $ \mathcal{W}^T / \mathcal{G}^T $.
This association is referred to as \emph{Berglund-H\"{u}bsch-Krawitz duality}.
Specifically, Krawitz established that
\begin{equation}
\mathcal{H}_{FJRW}(\mathcal{W},\mathcal{G}) 
   \cong \mathcal{Q} \left( \mathcal{W}^T,\mathcal{G}^T \right),
\end{equation}
where $ \mathcal{H}_{FJRW}(\mathcal{W},\mathcal{G}) $ is the 
\emph{Fan-Jarvis-Ruan-Witten state space} 
\cite{FanJarvisRuan:Geometry,FanJarvisRuan:TheWitten-mirror,FanJarvisRuan:TheWitten-virtual} of $ \mathcal{W} / \mathcal{G} $ and \\
$ \mathcal{Q} \left( \mathcal{W}^T,\mathcal{G}^T \right) $ is the 
\emph{orbifold Milnor ring} 
\cite{Krawitz:FJRW,IntriligatorVafa:Landau,Kaufmann:Singularities}
of $ \mathcal{W}^T / \mathcal{G}^T $.

Let $ \mathcal{W} = \sum_{i=1}^m c_i \prod_{j=1}^m \phi_j^{ a_{ij} } $ be an invertible nondegenerate quasihomogeneous polynomial Landau-Ginzburg potential.
The name \emph{invertible} means that the matrix $ \mathcal{A}_{\mathcal{W}} = (a_{ij}) $ of exponents is invertible.
Since $ \mathcal{A}_{\mathcal{W}} $ is invertible, the nonzero complex coefficients $ c_i $ may be absorbed by rescaling the $ \phi_j $.
Thus, without loss of generality, we can write
\begin{equation}
\label{W-invertible-nondegenerate}
\mathcal{W} = \sum_{i=1}^m  \prod_{j=1}^m \phi_j^{ a_{ij} } \, .
\end{equation}
A potential $ \mathcal{W}(\phi_1,\ldots,\phi_m) $ is \emph{nondegenerate} when its only critical point is at $ (\phi_1,\ldots,\phi_m) = (0,\ldots,0) $. 
We say that $ \mathcal{W}(\phi_1,\ldots,\phi_m) $ is a \emph{quasihomogeneous polynomial} of degree $ s $ when there exist positive integer weights 
$ n_{\phi_1},\ldots,n_{\phi_m} $ and a positive integer $ s $ such that
\begin{equation}
\begin{gathered}
\mathcal{W} \left( \lambda^{n_{\phi_1}} \phi_1, \ldots, \lambda^{n_{\phi_m}} \phi_m \right)
   = \lambda^s \mathcal{W} \left( \phi_1,\ldots,\phi_m \right)
\quad
\forall \, \lambda \in \mathbf{C} \, ,
\\
\gcd{(n_{\phi_1},\ldots,n_{\phi_m},s)} = 1 \, .
\end{gathered}
\end{equation}    

The group $ \mathrm{Aut}(\mathcal{W}) $ of diagonal automorphisms of $ \mathcal{W} $, i.e.
\begin{equation}
\mathrm{Aut}(\mathcal{W}):
   \
   (\phi_1,\ldots,\phi_m) \rightarrow (\omega_{\phi_1} \phi_1, \ldots, \omega_{\phi_m} \phi_m)
   \
   \left|
          \ 
          \prod_{j=1}^m \omega_{\phi_j}^{ a_{ij} } = 1 \, , 
          \quad
          i=1,\ldots,m \, ,
   \right.          
\end{equation}
contains two natural subgroups.
First, we have
\begin{multline}
\mathrm{SL}_{\mathcal{W}} 
   \equiv \mathrm{SL} \left( m,\mathbf{C} \right) \cap \mathrm{Aut}(\mathcal{W}) :
\\
   (\phi_1,\ldots,\phi_m) \rightarrow (\omega_{\phi_1} \phi_1, \ldots, \omega_{\phi_m} \phi_m)
   \
   \left|
          \            
          \begin{array}{l}
          \prod_{j=1}^m \omega_{\phi_j}^{ a_{ij} } = 1 \, , 
          \quad
          i=1,\ldots,m \, ;
          \\[1ex]
          \prod_{j=1}^m \omega_{\phi_j} = 1 \, .
          \end{array}
   \right. 
\end{multline}
Second, we have the cyclic group $ \langle J_{\mathcal{W}} \rangle $ of order $ s $ generated by
\begin{equation}
J_{\mathcal{W}}: 
   \
   (\phi_1,\ldots,\phi_m) 
   \rightarrow 
   \left(
          e^{2 \pi i \, n_{\phi_1} / s} \phi_1, \ldots, e^{2 \pi i \, n_{\phi_m} / s} \phi_m
   \right). 
\end{equation}
Krawitz \cite{Krawitz:FJRW} proved that any subgroup of $ \mathrm{Aut}(\mathcal{W}) $ containing $ J_{\mathcal{W}} $ is admissible.
That is, an admissible group $ \mathcal{G} $ satisfies
\begin{equation}
\langle J_{\mathcal{W}} \rangle \subset \mathcal{G} \subset \mathrm{Aut}(\mathcal{W}) \, .
\end{equation}

The dual potential potential $ \mathcal{W}^T $ is given by
\begin{equation}
\label{W^T-invertible-nondegenerate}
\mathcal{W}^T = \sum_{i=1}^m \prod_{j=1}^m y_j^{ a_{ji} } \, .
\end{equation}
It follows from \cite{KreuzerSkarke:On} that $ \mathcal{W}^T $ is an invertible nondegenerate quasihomogeneous polynomial of some degree $ s^{\prime} $.
We will denote the weights of the $ y_1,\ldots,y_m $ by $ n_{y_1},\ldots,n_{y_m} $, respectively.
The groups 
$ \mathrm{Aut}(\mathcal{W}^T) $, 
$ \mathrm{SL}_{\mathcal{W}^T} $, and  
$ \langle J_{\mathcal{W}^T} \rangle $
are defined in analogous way to which 
$ \mathrm{Aut}(\mathcal{W}) $, $ \mathrm{SL}_{\mathcal{W}} $, and  
$ \langle J_{\mathcal{W}} \rangle $ are defined, respectively.

A complete set of generators 
$ (\varrho_1,\ldots,\varrho_m) $ and
$ (\overline{\varrho}_1,\ldots,\overline{\varrho}_m) $ 
for the groups $ \mathrm{Aut}(\mathcal{W}) $ and $ \mathrm{Aut}(\mathcal{W}^T) $ can be read off from the columns and rows of $ \mathcal{A}_{\mathcal{W}}^{-1} $, respectively.
Let $ \rho_i = \left( \rho_i^{(1)},\ldots,\rho_i^{(m)} \right) $ be the $ i $-th column and
$ \overline{\rho}_i = \left( \overline{\rho}_i^{(1)},\ldots,\overline{\rho}_i^{(m)} \right) $
be the $ i $-th row of $ \mathcal{A}_{\mathcal{W}}^{-1} $.
Then
\begin{align*}
\varrho_i : 
  &\
   (\phi_1,\ldots,\phi_m) 
   \rightarrow 
   \left(
          e^{2 \pi i \, \rho_i^{(1)}} \phi_1, \ldots, e^{2 \pi i \, \rho_i^{(m)}} \phi_m
   \right),
\\[1ex]
\overline{\varrho}_i : 
  &\
   (y_1,\ldots,y_m) 
   \rightarrow 
   \left(
          e^{2 \pi i \, \overline{\rho}_i^{(1)}} y_1, \ldots, 
          e^{2 \pi i \, \overline{\rho}_i^{(m)}} y_m
   \right).     
\end{align*}
The dual group $ \mathcal{G}^T $ is given by
\begin{equation}
\mathcal{G}^T 
   = \left\{
             \prod_{i=1}^m \overline{\varrho}_i^{\beta_i} \, 
             \left| \,
                    \begin{bmatrix}
                       \beta_1 & \ldots & \beta_m
                    \end{bmatrix}
                    \mathcal{A}_{\mathcal{W}}^{-1}
                    \begin{bmatrix}
                       \alpha_1 \\ 
                       \vdots \\
                       \alpha_m
                    \end{bmatrix} 
                    \in \mathbf{Z}
                    \quad
                    \forall \prod_{i=1}^m \varrho_i^{\alpha_i} \in \mathcal{G}
             \right.
     \right\}.
\end{equation}

Chiodo and Ruan \cite{ChiodoRuan:LG/CY} established a Landau-Ginzburg/Calabi-Yau correspondence which allowed them to obtain a geometrical interpretation of Berglund-H\"{u}bsch-Krawitz duality when the zero locus of $ \mathcal{W} $ defines a Calabi-Yau hypersurface in a weighted complex projective space, i.e.
\begin{equation}
\mathcal{M}_{\mathcal{W}}   
   = \left\{ \mathcal{W} = 0 \right\}
   \in \mathbf{WCP}^{m-1}_{n_{\phi_1},\ldots,n_{\phi_m}}[s] \, ,   
\end{equation}
where
\begin{equation}
\label{CY-invertible-nondegenerate}
\sum_{i=1}^m n_{\phi_i} = s \, ,
\end{equation}
and the group $ \mathcal{G} $ satisfies
\begin{equation}
\langle J_{\mathcal{W}} \rangle
   \subset \mathcal{G}  
   \subset \mathrm{SL}_{\mathcal{W}} \, . 
\end{equation}
Under these assumptions, the following properties hold:
\begin{enumerate}

\item
The zero locus of $ \mathcal{W}^T $ defines a Calabi-Yau hypersurface in a weighted complex projective space, i.e.
\begin{equation}
\mathcal{M}_{\mathcal{W}^T}   
   = \left\{ \mathcal{W}^T = 0 \right\}
   \in \mathbf{WCP}^{m-1}_{n_{y_1},\ldots,n_{y_m}}[s^{\prime}] \, ,   
\end{equation}
where
\begin{equation}
\sum_{i=1}^m n_{y_i} = s^{\prime} \, .
\end{equation}

\item
The group $ \mathcal{G}^T $ satisfies
\begin{equation}
\langle J_{\mathcal{W}^T} \rangle
   \subset \mathcal{G}^T  
   \subset \mathrm{SL}_{\mathcal{W}^T} \, . 
\end{equation}

\item
The Calabi-Yau orbifolds 
$ 
\mathcal{M}_{\mathcal{W}} /
\left( 
       \mathcal{G} / 
       \langle J_{\mathcal{W}} \rangle  
\right) 
$  
and 
$ 
\mathcal{M}_{\mathcal{W}^T} /
\left( 
       \mathcal{G}^T / 
       \langle J_{\mathcal{W}^T} \rangle  
\right)
$
form a mirror pair, i.e.
\begin{equation}
h^{p,q}_{CR}
\left(
       \frac{ \mathcal{M}_{\mathcal{W}} }
            { \mathcal{G} / \langle J_{\mathcal{W}} \rangle }  
\right)
  = h^{m-2-p,q}_{CR}
    \left(
           \frac{ \mathcal{M}_{\mathcal{W}^T} }
                { \mathcal{G}^T / \langle J_{\mathcal{W}^T} \rangle }  
   \right) \, ,
\end{equation}
where $ h^{p,q}_{CR} $ is the \emph{Chen-Ruan orbifold cohomology} \cite{ChenRuan:Anew}.

\end{enumerate}

\subsection{Hori-Vafa point of view}

We will now discuss the Chiodo-Ruan geometrical interpretation of 
Berglund-H\"{u}bsch-Krawitz duality in the Hori-Vafa context.
Setting $ k = 1 $ in  (\ref{mirror-period-M_G}) yields
\begin{align}
\Pi_{\widetilde{\mathcal{M}}_G} 
  &= \int       
     \left(
            \prod_{i=1}^m dY_i
     \right)     
     dY_P \, e^{-Y_P} \,   
     \delta \left(
                   \sum_{i=1}^m Q_i Y_i - s Y_P - t
            \right)
     \exp{ \left(
              - \sum_{i=1}^m  e^{-Y_i} - e^{-Y_P}
        \right)
         } 
\nonumber
\\[1ex]
  &= \int       
     \left(
            \prod_{i=1}^m dY_i
     \right)
     \left[
             e^{t/s} \prod_{i=1}^m \left( e^{-Y_i} \right)^{Q_i / s}       
      \right]
      \exp{ \left[
                 - \sum_{i=1}^m  e^{-Y_i} 
                 - e^{t/s} 
                   \prod_{i=1}^m  \left( e^{-Y_i} \right)^{Q_i /s}
            \right]
          }. 
\end{align}
Suppose there exists an invertible matrix $ (M_{ji}) $ of nonnegative integers such that 
\begin{equation} 
\label{M-equations:WCP^{m-1}[s];k=1} 
s = \sum_{i=1}^m M_{ji} Q_i \, , 
\qquad  
j = 1,\ldots,m \, . 
\end{equation} 
Now, consider the change of variables  
\begin{gather}
\label{change-of-variables} 
e^{-Y_i} = \prod_{j=1}^m y_j^{ M_{j i} } \, . 
\end{gather} 
This change of variables one-to-one up to the action of the group $ \Gamma $ defined by 
\begin{equation} 
\label{Gamma:WCP^{m-1}[s];k=1} 
\Gamma : 
\quad 
(y_1,\ldots,y_m) \rightarrow (\omega_{y_1} y_1,\ldots,\omega_{y_m} y_m)
\quad 
\left | 
        \quad
        \begin{array}{l}
        \prod_{j=1}^m \omega_{y_j}^{M_{ji}} = 1 \, ,
        \qquad 
        i = 1,\ldots,m \, ;  
        \\[1ex]
        \prod_{j=1}^m \omega_{y_j} = 1 \, .
        \end{array} 
\right.
\end{equation}
In terms of the new variables, we obtain 
\begin{align} 
\Pi_{\widetilde{\mathcal{M}}_G} 
  = (-1)^m \det{(M_{ji})}  \, e^{t/s} \int 
    \left( 
           \prod_{i=1}^m dy_i 
    \right)          
    \exp{ \left[ 
               - \sum_{i=1}^m  \prod_{j=1}^m y_j^{ M_{ji} } 
               - e^{t/s} \prod_{j=1}^m y_j            
         \right] 
        }. 
\end{align} 
This is the period for the Landau-Ginzburg orbifold   
$ \widetilde{W} / \Gamma \, , $ where  
\begin{equation} 
\label{Wtilde:WCP^{m-1}[s];k=1} 
\widetilde{W}  
   = \sum_{i=1}^m  \prod_{j=1}^m y_j^{ M_{ji} } 
    + e^{t/s} \prod_{j=1}^m y_j \, .
\end{equation} 
As proven in \cite{GaravusoKatzarkovKreuzerNoll}, $ \widetilde{W} $ is quasihomogeneous of some degree $ s^{\prime} $ for all values of $ t $ if and only if the Calabi-Yau condition $ \sum_{i=1}^m Q_i  = s $ is satisfied.

Now, set $ G = \mathcal{W} $, where $ \mathcal{W} $ is given by 
(\ref{W-invertible-nondegenerate}).
Taking the limit $ t \rightarrow - \infty $ and setting $ M_{ji} = a_{ji} $ in
(\ref{Wtilde:WCP^{m-1}[s];k=1}) yields the expression for $ \mathcal{W}^T $ given by 
(\ref{W^T-invertible-nondegenerate}).
We thus obtain the Landau-Ginzburg orbifold $ \mathcal{W}^T / \mathrm{SL}_{\mathcal{W}^T} $.
Imposing the Calabi-Yau condition (\ref{CY-invertible-nondegenerate}) with
$ (n_{\phi_1},\ldots,n_{\phi_m} ) = (Q_1,\ldots,Q_m) $, we obtain the mirror pair
$ ( \mathcal{M}_{\mathcal{W}}, \widetilde{\mathcal{M}}_{\mathcal{W}} ) $, where
\begin{align}
\label{M_W-invertible-nondegenerate}
\mathcal{M}_{\mathcal{W}}   
  &= \left\{ \mathcal{W} = 0 \right\}
   \in \mathbf{WCP}^{m-1}_{n_{\phi_1},\ldots,n_{\phi_m}}[s] \, , 
  &\sum_{i=1}^m n_{\phi_i} &= s \, , 
\\[1ex]
\label{M_W^T-invertible-nondegenerate}
\widetilde{\mathcal{M}}_{\mathcal{W}}   
  &= \frac{ 
            \left\{ \mathcal{W}^T = 0 \right\}
            \in \mathbf{WCP}^{m-1}_{n_{y_1},\ldots,n_{y_m}}[s^{\prime}] 
          }
          { \mathrm{SL}_{\mathcal{W}^T} / \langle J_{\mathcal{W}^T} \rangle } \, , 
  &\sum_{i=1}^m n_{y_i} &= s^{\prime} \, .          
\end{align}

\subsection{Application to \texorpdfstring{$\mathbf{CP}^2[3]$}{CP-2[3]},
                           \texorpdfstring{$\mathbf{CP}^3[4]$}{CP-3[4]},
                       and \texorpdfstring{$\mathbf{CP}^4[5]$}{CP-4[5]}}
                       
Tables \ref{Table-CP^2[3]}, \ref{Table-CP^3[4]}, and \ref{Table-CP^4[5]} 
list mirror pairs 
$ \left( \mathcal{M}_{\mathcal{W}}, \widetilde{\mathcal{M}}_{\mathcal{W}} \right) $ given by 
(\ref{M_W-invertible-nondegenerate}) and 
(\ref{M_W^T-invertible-nondegenerate}) when 
$ \mathcal{M}_{\mathcal{W}} \in \mathbf{CP}^2[3] $, 
$ \mathcal{M}_{\mathcal{W}} \in \mathbf{CP}^3[4] $, and
$ \mathcal{M}_{\mathcal{W}} \in \mathbf{CP}^4[5] $, respectively. 
These tables are complete in the sense that \emph{all} inequivalent invertible nondegenerate quasihomogeneous potentials $ \mathcal{W} $ appropriate to each table are considered. 
When the orbifold group $ \mathrm{SL}_{\mathcal{W}^T} / \langle J_{\mathcal{W}^T} \rangle $ is nontrivial, we use the shorthand $ \mathbf{Z}_k: \ [r_1,\ldots,r_m] $ to denote a 
$ \mathbf{Z}_k $ symmetry with action
$ (y_1,\ldots,y_m) \rightarrow (\alpha^{r_1} y_1,\ldots,\alpha^{r_m} y_m) $, where 
$ \alpha^k = 1 $. 

\begin{table}
\tiny
\begin{center}
$  
\begin{tabular}{r|c|c|}
\cline{2-3}
& &
\\[-1ex] &
$ \mathcal{M}_{\mathcal{W}} = \left\{ \mathcal{W} = 0 \right\} $  & 
$ {\displaystyle 
  \widetilde{\mathcal{M}}_{\mathcal{W}} 
     = \frac{ \left\{ \mathcal{W}^T = 0 \right\} }
            { \mathrm{SL}_{\mathcal{W}^T} / \langle J_{\mathcal{W}^T} \rangle } 
  } $
\\[-1ex]
& &
\\
\cline{2-3}
& &
\\ 1 &
$ \left\{ \phi_1^3 + \phi_2^3 + \phi_3^3  = 0 \right\} 
  \in \mathbf{CP}^2[3] $ &  
$ {\displaystyle
  \frac{ \left\{ y_1^3 + y_2^3 + y_3^3 = 0 \right\} \in \mathbf{CP}^2[3] }
       { \mathbf{Z}_3 : \ [2,1,0] } 
  } $         
\\
& &
\\
\cline{2-3}
& &
\\ 2 &
$ \left\{ \phi_1^2 \phi_2 + \phi_2^3 + \phi_3^3 = 0 \right\} 
  \in \mathbf{CP}^2[3] $ &
$ {\displaystyle
  \left\{ y_1^2 + y_1 y_2^3 + y_3^3 = 0 \right\} 
  \in \mathbf{WCP}^2_{3,1,2}[6] 
  } $
\\
& &
\\
\cline{2-3}
& &
\\ 3 &
$ \left\{ \phi_1^2 \phi_2 + \phi_2^2 \phi_3 + \phi_3^3 = 0 \right\} 
  \in \mathbf{CP}^2[3] $ &
$ {\displaystyle 
  \left\{ y_1^2 + y_1 y_2^2 + y_2 y_3^3  = 0 \right\} 
  \in \mathbf{WCP}^2_{2,1,1}[4]
  } $
\\
& &
\\
\cline{2-3}
& &
\\ 4 &
$ \left\{ \phi_1^3 + \phi_2^2 \phi_3 + \phi_2 \phi_3^2 = 0 \right\} 
  \in \mathbf{CP}^2[3] $ & 
$ {\displaystyle
  \left\{ y_1^3 + y_2^2 y_3 + y_2 y_3^2 = 0 \right\} 
  \in \mathbf{CP}^2[3]
  } $
\\
& &
\\
\cline{2-3}
& &
\\ 5 &
$ \left\{ \phi_1^2 \phi_2 + \phi_2^2 \phi_3 + \phi_1 \phi_3^2 = 0 \right\} 
  \in \mathbf{CP}^2[3] $ & 
$ {\displaystyle
  \left\{  y_1^2 y_3 + y_1 y_2^2 + y_2 y_3^2 = 0 \right\} 
  \in \mathbf{CP}^2[3]
  } $ 
\\
& &
\\
\cline{2-3}
\end{tabular}
$
\caption{
Mirror pairs 
$ \left( \mathcal{M}_{\mathcal{W}}, \widetilde{\mathcal{M}}_{\mathcal{W}} \right) $
when $ \mathcal{M}_{\mathcal{W}} \in \mathbf{CP}^2[3] $.
        }
\label{Table-CP^2[3]}
\end{center}
\end{table}

\begin{table}[H]
\tiny
\begin{center}
$  
\begin{tabular}{r|c|c|}
\cline{2-3}
& &
\\[-1ex] &
$ \mathcal{M}_{\mathcal{W}} = \left\{ \mathcal{W} = 0 \right\} $  & 
$ {\displaystyle 
  \widetilde{\mathcal{M}}_{\mathcal{W}} 
     = \frac{ \left\{ \mathcal{\mathcal{W}}^T = 0 \right\} }
            { \mathrm{SL}_{\mathcal{W}^T} / \langle J_{\mathcal{W}^T} \rangle } 
  } $
\\[-1ex]
& &
\\
\cline{2-3}
& &
\\ 1 &
$ \left\{ \phi_1^4 + \phi_2^4 + \phi_3^4 + \phi_4^4 = 0 \right\} 
  \in \mathbf{CP}^3[4] $ &  
$ {\displaystyle
  \frac{ \left\{ y_1^4 + y_2^4 + y_3^4 + y_4^4 = 0 \right\} \in \mathbf{CP}^3[4] }
       { (\mathbf{Z}_4)^2 : \ [3,1,0,0] \, , \, [3,0,1,0] } 
  } $         
\\
& &
\\
\cline{2-3}
& &
\\ 2 &
$ \left\{ \phi_1^3 \phi_2 + \phi_2^4 + \phi_3^4 + \phi_4^4 = 0 \right\} 
  \in \mathbf{CP}^3[4] $ &
$ {\displaystyle
  \frac{ \left\{ y_1^3 + y_1 y_2^4 + y_3^4 + y_4^4 = 0 \right\} 
         \in \mathbf{WCP}^3_{4,2,3,3}[12] }
       { \mathbf{Z}_4 : \ [0,0,3,1] }
  } $
\\
& &
\\
\cline{2-3}
& &
\\ 3 &
$ \left\{ \phi_1^3 \phi_2 + \phi_2^3 \phi_3 + \phi_3^4 + \phi_4^4 = 0 \right\} 
  \in \mathbf{CP}^3[4] $ &
$ {\displaystyle 
  \left\{ y_1^3 + y_1 y_2^3 + y_2 y_3^4 + y_4^4 = 0 \right\} 
  \in \mathbf{WCP}^3_{12,8,7,9}[36]
  } $
\\
& &
\\
\cline{2-3}
& &
\\ 4 &
$ \left\{ \phi_1^3 \phi_2 + \phi_2^3 \phi_3 + \phi_3^3 \phi_4  + \phi_4^4 = 0 \right\} 
  \in \mathbf{CP}^3[4] $ &
$ {\displaystyle 
  \left\{ y_1^3 + y_1 y_2^3 + y_2 y_3^3 + y_3 y_4^4 = 0 \right\} 
  \in \mathbf{WCP}^3_{9,6,7,5}[27]
  } $
\\
& &
\\
\cline{2-3}
& &
\\ 5 &
$ \left\{ \phi_1^3 \phi_2 + \phi_2^4 + \phi_3^3 \phi_4 + \phi_4^4 = 0 \right\} 
  \in \mathbf{CP}^3[4] $ &
$ {\displaystyle
  \frac{ \left\{ y_1^3 + y_1 y_2^4 + y_3^3 + y_3 y_4^4 = 0 \right\} 
         \in \mathbf{WCP}^3_{2,1,2,1}[6] }
       { \mathbf{Z}_2 : \ [0,1,0,1] }  
  } $
\\
& &
\\
\cline{2-3}
& &
\\ 6 &
$ \left\{ \phi_1^4 + \phi_2^4  + \phi_3^3 \phi_4 + \phi_3 \phi_4^3 = 0 \right\}
  \in \mathbf{CP}^3[4] $ &
$ {\displaystyle 
  \frac{ \left\{ y_1^4 + y_2^4 + y_3^3 y_4 + y_3 y_4^3 = 0 \right\} 
         \in \mathbf{CP}^3[4] }
       { \mathbf{Z}_8 : \ [0,2,5,1] }
  } $
\\
& &
\\
\cline{2-3}
& &
\\ 7 &
$ \left\{ \phi_1^3 \phi_2 + \phi_2^4 + \phi_3^3 \phi_4 + \phi_3 \phi_4^3 = 0 \right\} 
  \in \mathbf{CP}^3[4] $ &
$ {\displaystyle
  \frac{ \left\{ y_1^3 + y_1 y_2^4 + y_3^3 y_4 + y_3 y_4^3 = 0 \right\} 
  \in \mathbf{WCP}^3_{4,2,3,3}[12] }
  { \mathbf{Z}_8 : \ [0,2,5,1] }
  } $
\\
& &
\\
\cline{2-3}
& &
\\ 8 &
$ \left\{ \phi_1^3 \phi_2 + \phi_1 \phi_2^3 + \phi_3^3 \phi_4 + \phi_3 \phi_4^3 = 0 \right\} 
  \in \mathbf{CP}^3[4] $ & 
$ {\displaystyle
  \frac{ \left\{ y_1^3 y_2 + y_1 y_2^3 + y_3^3 y_4 + y_3 y_4^3 = 0 \right\} 
         \in \mathbf{CP}^3[4] }
      { \mathbf{Z}_2 : \ [1,1,1,1] }
  } $
\\
& &
\\
\cline{2-3}
& &
\\ 9 &
$ \left\{ \phi_1^4 + \phi_2^3 \phi_3 + \phi_3^3 \phi_4 + \phi_2 \phi_4^3 = 0 \right\} 
  \in \mathbf{CP}^3[4] $ &
$ {\displaystyle
  \frac{ \left\{ y_1^4 + y_2^3 y_4 + y_2 y_3^3 + y_3 y_4^3 = 0 \right\} \in \mathbf{CP}^3[4] }
       { \mathbf{Z}_7 : \ [0,2,4,1] }
  } $
\\
& &
\\
\cline{2-3}
& &
\\ 10 &
$ \left\{ \phi_1^3 \phi_2 + \phi_2^3 \phi_3 + \phi_3^3 \phi_4 + \phi_1 \phi_4^3 = 0 \right\} 
  \in \mathbf{CP}^3[4] $ & 
$ {\displaystyle
  \frac{ \left\{  y_1^3 y_4 + y_1 y_2^3 + y_2 y_3^3 + y_3 y_4^3 = 0 \right\} \in \mathbf{CP}^3[4] }
       { \mathbf{Z}_5 : \ [3,4,2,1] }
  } $ 
\\
& &
\\
\cline{2-3}
\end{tabular}
$
\caption{
Mirror pairs 
$ \left( \mathcal{M}_{\mathcal{W}}, \widetilde{\mathcal{M}}_{\mathcal{W}} \right) $
when $ \mathcal{M}_{\mathcal{W}} \in \mathbf{CP}^3[4] $.
        }
\label{Table-CP^3[4]}
\end{center}
\end{table}

\begin{table}[H]
\tiny
\setlength{\tabcolsep}{5pt}
\begin{center}
$  
\begin{tabular}{r|c|c|}
\cline{2-3}
& &
\\[-1ex] &
$ \mathcal{M}_{\mathcal{W}} = \left\{ \mathcal{W} = 0 \right\} $  & 
$ {\displaystyle 
  \widetilde{\mathcal{M}}_{\mathcal{W}} 
     = \frac{ \left\{ \mathcal{W}^T = 0 \right\} }
            { \mathrm{SL}_{\mathcal{W}^T} / \langle J_{\mathcal{W}^T} \rangle } 
  } $
\\[-1ex]
& &
\\
\cline{2-3}
& &
\\ 1 &
$ \left\{ \phi_1^5 + \phi_2^5 + \phi_3^5 + \phi_4^5 + \phi_5^5 = 0 \right\} 
  \in \mathbf{CP}^4[5] $ &  
$ {\displaystyle
  \frac{ \left\{ y_1^5 + y_2^5 + y_3^5 + y_4^5 + y_5^5  = 0 \right\} \in \mathbf{CP}^4[5] }
       { (\mathbf{Z}_5)^3 : \ [4,1,0,0,0] \, , \, [4,0,1,0,0] \, , \, [4,0,0,1,0] } 
  } $         
\\
& &
\\
\cline{2-3}
& &
\\ 2 &
$ \left\{ \phi_1^4 \phi_2 + \phi_2^5 + \phi_3^5 + \phi_4^5 + \phi_5^5 = 0 \right\} 
  \in \mathbf{CP}^4[5] $ &
$ {\displaystyle
  \frac{ \left\{ y_1^4 + y_1 y_2^5 + y_3^5 + y_4^5 + y_5^5 = 0 \right\} 
         \in \mathbf{WCP}^4_{5,3,4,4,4}[20] }
       { (\mathbf{Z}_5)^2 : \ [0,0,4,1,0] \, , \, [0,0,4,0,1] }
  } $
\\
& &
\\
\cline{2-3}
& &
\\ 3 &
$ \left\{ \phi_1^4 \phi_2 + \phi_2^4 \phi_3 + \phi_3^5 + \phi_4^5 + \phi_5^5 = 0 \right\} 
  \in \mathbf{CP}^4[5] $ &
$ {\displaystyle 
  \frac{ \left\{ y_1^4 + y_1 y_2^4 + y_2 y_3^5 + y_4^5 + y_5^5 = 0 \right\} 
         \in \mathbf{WCP}^4_{20,15,13,16,16}[80] }
       { \mathbf{Z}_5 : \ [0,0,0,4,1] } 
  } $
\\
& &
\\
\cline{2-3}
& &
\\ 4 & 
$ \left\{ \phi_1^4 \phi_2 + \phi_2^4 \phi_3 + \phi_3^4 \phi_4  + \phi_4^5 + \phi_5^5 = 0 \right\} 
  \in \mathbf{CP}^4[5] $ &
$ {\displaystyle 
  \left\{ y_1^4 + y_1 y_2^4 + y_2 y_3^4 + y_3 y_4^5 + y_5^5 = 0 \right\} 
  \in \mathbf{WCP}^4_{80,60,65,51,64}[320]
  } $
\\
& &
\\
\cline{2-3}
& &
\\ 5 &
$ \left\{ \phi_1^4 \phi_2 + \phi_2^4 \phi_3 + \phi_3^4 \phi_4  + \phi_4^4 \phi_5 + \phi_5^5 = 0 \right\} 
  \in \mathbf{CP}^4[5] $ &
$ {\displaystyle
  \left\{ y_1^4 + y_1 y_2^4 + y_2 y_3^4 + y_3 y_4^4 + y_4 y_5^5 = 0 \right\} 
  \in \mathbf{WCP}^4_{64,48,52,51,41}[256]
  } $
\\
& &
\\
\cline{2-3}
& &
\\ 6 &
$ \left\{ \phi_1^4 \phi_2 + \phi_2^5 + \phi_3^4 \phi_4 + \phi_4^5 + \phi_5^5 = 0 \right\} 
  \in \mathbf{CP}^4[5] $ &
$ {\displaystyle 
  \left\{ y_1^4 + y_1 y_2^5 + y_3^4 + y_3 y_4^5 + y_5^5 = 0 \right\} 
  \in \mathbf{WCP}^4_{5,3,5,3,4}[20]
  } $
\\
& &
\\
\cline{2-3}
& &
\\ 7 &
$ \left\{ \phi_1^4 \phi_2 + \phi_2^4 \phi_3 + \phi_3^5 + \phi_4^4 \phi_5 + \phi_5^5 = 0 \right\}
  \in \mathbf{CP}^4[5] $ &
$ {\displaystyle
  \left\{ y_1^4 + y_1 y_2^4 + y_2 y_3^5 + y_4^4 + y_4 y_5^5  = 0 \right\} 
  \in \mathbf{WCP}^4_{20,15,13,20,12}[80]
  } $
\\
& &
\\
\cline{2-3}
& &
\\ 8 &
$ \left\{ \phi_1^5 + \phi_2^5 + \phi_3^5 + \phi_4^4 \phi_5 + \phi_4 \phi_5^4 = 0 \right\}
  \in \mathbf{CP}^4[5] $ &
$ {\displaystyle 
  \frac{ \left\{ y_1^5 + y_2^5 + y_3^5 + y_4^4 y_5 + y_4 y_5^4 = 0 \right\} 
         \in \mathbf{CP}^4[5] }
       { (\mathbf{Z}_5)^2 \times \mathbf{Z}_3 : \ [1,4,0,0,0] \, , \, [0,4,1,0,0] \, , \, [0,0,0,2,1] }
  } $
\\
& &
\\
\cline{2-3}
& &
\\ 9 &
$ \left\{ \phi_1^5 + \phi_2^4 \phi_3 + \phi_3^5 + \phi_4^4 \phi_5 + \phi_4 \phi_5^4 = 0 \right\} 
  \in \mathbf{CP}^4[5] $ &
$ {\displaystyle
  \left\{ y_1^5 + y_2^4 + y_2 y_3^5 + y_4^4 y_5 + y_4 y_5^4 = 0 \right\} 
  \in \mathbf{WCP}^4_{4,5,3,4,4}[20] 
  } $
\\
& &
\\
\cline{2-3}
& &
\\ 10 &
$ \left\{ \phi_1^4 \phi_2 + \phi_2^4 \phi_3 + \phi_3^5 + \phi_4^4 \phi_5 + \phi_4 \phi_5^4 = 0 \right\} 
  \in \mathbf{CP}^4[5] $ &
$ {\displaystyle
  \frac{ \left\{ y_1^4 + y_1 y_2^4 + y_2 y_3^5 + y_4^4 y_5 + y_4 y_5^4 = 0 \right\} 
         \in \mathbf{WCP}^4_{20,15,13,16,16}[80] } 
       { \mathbf{Z}_{15} : \ [0,0,3,11,1] }
  } $
\\
& &
\\
\cline{2-3}
& &
\\ 11 & 
$ \left\{ \phi_1^5 + \phi_2^4 \phi_3 + \phi_2 \phi_3^4 + \phi_4^4 \phi_5 + \phi_4 \phi_5^4 = 0 \right\} 
  \in \mathbf{CP}^4[5] $ & 
$ {\displaystyle
  \frac{ \left\{ y_1^5 + y_2^4 y_3 + y_2 y_3^4 + y_4^4 y_5 + y_4 y_5^4 = 0 \right\} 
         \in \mathbf{CP}^4[5] }
      { \mathbf{Z_3} : \ [0,2,1,2,1] }
  } $
\\
& &
\\
\cline{2-3}
& &
\\ 12 &
$ \left\{ \phi_1^5 + \phi_2^5 + \phi_3^4 \phi_4 + \phi_4^4 \phi_5 + \phi_3 \phi_5^4 = 0 \right\} 
  \in \mathbf{CP}^4[5] $ &
$ {\displaystyle
  \frac{ \left\{ y_1^5 + y_2^5 + y_3^4 y_5 + y_3 y_4^4 + y_4 y_5^4 = 0 \right\} \in \mathbf{CP}^4[5] }
       { \mathbf{Z}_5 \times \mathbf{Z}_{13} : \ [1,4,0,0,0] \, , \, [0,0,3,9,1] }
  } $
\\
& &
\\
\cline{2-3}
& &
\\ 13 &
$ \left\{ \phi_1^4 \phi_2 + \phi_2^5 + \phi_3^4 \phi_4 + \phi_4^4 \phi_5 + \phi_3 \phi_5^4 = 0 \right\} 
  \in \mathbf{CP}^4[5] $ &
$ {\displaystyle
  \frac{ \left\{ y_1^4 + y_1 y_2^5 + y_3^4 y_5 + y_3 y_4^4 + y_4 y_5^4 = 0 \right\} 
         \in \mathbf{WCP}^4_{5,3,4,4,4}[20] }
       { \mathbf{Z}_{65} : \ [0,52,16,61,1] }
  } $ 
\\
& &
\\
\cline{2-3}
& &
\\ 14 &
$ \left\{ \phi_1^4 \phi_2 + \phi_1 \phi_2^4 + \phi_3^4 \phi_4 + \phi_4^4 \phi_5 + \phi_3 \phi_5^4 = 0 \right\} 
  \in \mathbf{CP}^4[5] $ &
$ {\displaystyle
  \frac{ \left\{ y_1^4 y_2 + y_1 y_2^4 + y_3^4 y_5 + y_3 y_4^4 + y_4 y_5^4 = 0 \right\} 
         \in \mathbf{CP}^4[5] }
       { \mathbf{Z}_3 \times \mathbf{Z}_{13} : \ [1,2,0,0,0] \, , \, [0,0,3,9,1] } 
  } $
\\
& &
\\
\cline{2-3}
& &
\\ 15 &
$ \left\{ \phi_1^5 + \phi_2^4 \phi_3 + \phi_3^4 \phi_4 + \phi_4^4 \phi_5 + \phi_2 \phi_5^4 = 0 \right\} 
  \in \mathbf{CP}^4[5] $ &
$ {\displaystyle
  \frac{ \left\{ y_1^5 + y_2^4 y_5 + y_2 y_3^4 + y_3 y_4^4 + y_4 y_5^4 = 0 \right\} 
         \in \mathbf{CP}^4[5] }
       { \mathbf{Z}_{51} : \ [0,38,16,47,1] } 
  } $
\\
& &
\\
\cline{2-3}
& &
\\ 16 &
$ \left\{ \phi_1^4 \phi_2 + \phi_2^4 \phi_3 + \phi_3^4 \phi_4 + \phi_4^4 \phi_5 + \phi_1 \phi_5^4 = 0 \right\} 
  \in \mathbf{CP}^4[5] $ & 
$ {\displaystyle
  \frac{ \left\{  y_1^4 y_5 + y_1 y_2^4 + y_2 y_3^4 + y_3 y_4^4 + y_4 y_5^4 = 0 \right\} \in \mathbf{CP}^4[5] }
       { \mathbf{Z}_{41} : \ [10,18,16,37,1] }
  } $ 
\\
& &
\\
\cline{2-3}
\end{tabular}
$
\caption{
Mirror pairs 
$ \left( \mathcal{M}_{\mathcal{W}}, \widetilde{\mathcal{M}}_{\mathcal{W}} \right) $
when $ \mathcal{M}_{\mathcal{W}} \in \mathbf{CP}^4[5] $.
        }
\label{Table-CP^4[5]}
\end{center}
\end{table}

Note that rows 1, 8, 14, 15, and 16 of Table \ref{Table-CP^4[5]} agree with the corresponding results presented in \cite{GreenePlesser:Mirror}.
The following example illustrates the calculations involved in obtaining Tables 
\ref{Table-CP^2[3]}, \ref{Table-CP^3[4]}, and \ref{Table-CP^4[5]}.
\begin{example}
Consider row 13 of Table \ref{Table-CP^4[5]}.
The potential $ \mathcal{W} $ is
\begin{equation*}
\mathcal{W} 
   = \sum_{i=1}^5 \prod_{j=1}^5 \phi_j^{a_{ij}}
   = \phi_1^4 \phi_2 + \phi_2^5 + \phi_3^4 \phi_4 + \phi_4^4 \phi_5 + \phi_3 \phi_5^4  \, .
\end{equation*}
From this expression, we obtain the matrix
\begin{equation*}
(a_{ij})
   = \begin{pmatrix}
        4 & 1 & 0 & 0 & 0 \\
        0 & 5 & 0 & 0 & 0 \\
        0 & 0 & 4 & 1 & 0 \\
        0 & 0 & 0 & 4 & 1 \\
        0 & 0 & 1 & 0 & 4  
     \end{pmatrix}  
\end{equation*}
and the dual potential
\begin{equation*}
\mathcal{W}^{T}
   = \sum_{i=1}^5 \prod_{j=1}^5 y_j^{a_{ji}}
   = y_1^4 + y_1 y_2^5 + y_3^4 y_5 + y_3 y_4^4 + y_4 y_5^4 \, .   
\end{equation*}
Thus,
\begin{align*}
\mathcal{M}_{\mathcal{W}}
  &= \left\{ \mathcal{W} = 0 \right\} \in \mathbf{CP}^4[5] \, ,
\\[1ex]
\mathcal{M}_{\mathcal{W}^T}
  &= \left\{ \mathcal{W}^T = 0 \right\} \in \mathbf{WCP}^4_{5,3,4,4,4}[20] \, .   
\end{align*}
Now, let us determine the orbifold group
$ \mathrm{SL}_{\mathcal{W}^T} / \langle J_{\mathcal{W}^T} \rangle $.
The action of $ \mathrm{SL}_{\mathcal{W}^T} $ is
\begin{equation*}  
\mathrm{SL}_{\mathcal{W}^T} : 
\quad 
(y_1,\ldots,y_5) \rightarrow (\omega_{y_1} y_1,\ldots,\omega_{y_5} y_5)
\quad 
\left | 
        \quad
        \begin{array}{l}
        1 = \prod_{j=1}^5 \omega_{y_j}^{a_{j1}} = \omega_{y_1}^4 \, ,
        \\[1ex]
        1 = \prod_{j=1}^5 \omega_{y_j}^{a_{j2}} = \omega_{y_1} \omega_{y_2}^5 \, ,
        \\[1ex]
        1 = \prod_{j=1}^5 \omega_{y_j}^{a_{j3}} = \omega_{y_3}^4 \omega_{y_5} \, ,
        \\[1ex]
        1 = \prod_{j=1}^5 \omega_{y_j}^{a_{j4}} = \omega_{y_3} \omega_{y_4}^4 \, ,
        \\[1ex]
        1 = \prod_{j=1}^5 \omega_{y_j}^{a_{j5}} =  \omega_{y_4} \omega_{y_5}^4 \, ,
        \\[1ex]
        1 = \prod_{j=1}^5 \omega_{y_j} \, .
        \end{array} 
\right.
\end{equation*}
Combining the second, fourth, and fifth constraints yields
\begin{equation*}
\left( 
       \omega_{y_1}, \, \omega_{y_2}, \, \omega_{y_3}, \, \omega_{y_4}, \, \omega_{y_5} 
\right) 
   = \left( 
            \omega_{y_2}^{-5}, \, \omega_{y_2}, \, \omega_{y_5}^{16}, \, 
            \omega_{y_5}^{-4}, \, \omega_{y_5}
     \right).
\end{equation*}
Imposing the sixth constraint on this result gives
\begin{equation*}
\omega_{y_2}^{-4} \omega_{y_5}^{13} = 1 \, .
\end{equation*}
It follows that
\begin{equation*}
\left( 
       \omega_{y_1}, \, \omega_{y_2}, \, \omega_{y_3}, \, \omega_{y_4}, \, \omega_{y_5} 
\right) 
   = \left( 
            \omega_{y_2}^{-5}, \, \omega_{y_2}^5 \omega_{y_5}^{-13}, \, 
            \omega_{y_5}^{16}, \, \omega_{y_5}^{-4}, \, \omega_{y_5}
     \right).
\end{equation*}
From the first and second constraints we obtain
\begin{equation*}
\omega_{y_2}^{20} = 1 \, ,
\end{equation*}
which together with our result $ \omega_{y_2}^{-4} \omega_{y_5}^{13} = 1 $ implies
\begin{equation*}
\omega_{y_5}^{65} = 1 \, .
\end{equation*}
We also obtain $ \omega_{y_5}^{65} = 1 $ by combining our result 
$ \omega_{y_3} = \omega_{y_5}^{16} $ with the third constraint. 
Thus, all six constraints have been satisfied and we obtain 
\begin{equation*}
\mathrm{SL}_{\mathcal{W}^T}
   = \mathbf{Z}_{20} \times Z_{65} : \ [15,5,0,0,0] \, , \, [0,52,16,61,1] \, .
\end{equation*}
Modding out by $ \langle J_{\mathcal{W}^T} \rangle  = \mathbf{Z}_{20} $ gives
\begin{equation*}
\mathrm{SL}_{\mathcal{W}^T} / \langle J_{\mathcal{W}^T} \rangle
   = Z_{65} : \ [0,52,16,61,1] \, .
\end{equation*}
We conclude that
\begin{equation*}
\widetilde{\mathcal{M}}_{\mathcal{W}}
   = \frac{ \left\{ \mathcal{W}^{T} = 0 \right\} \in \mathbf{WCP}^4_{5,3,4,4,4}[20] }
          { \mathbf{Z}_{65} : \ [0,52,16,61,1] } \, .
\end{equation*}
\end{example}

In Tables \ref{Table-CP^2[3]}, \ref{Table-CP^3[4]}, and \ref{Table-CP^4[5]}, the left columns correspond to different complex structures for $ \mathcal{M}_{\mathcal{W}} $ whereas the right columns correspond to different K\"{a}hler structures for 
$ \widetilde{\mathcal{M}}_{\mathcal{W}} $.
We can probe the different K\"{a}hler structures for $ \widetilde{\mathcal{M}}_{\mathcal{W}} $
by computing the associated Picard-Fuchs equations. 
Doing this for rows 2 and 3 of Table \ref{Table-CP^2[3]}, we obtain
\begin{align}
0 &= \left[ 
            \theta^2 - 12 e^{-t} (6 \theta + 5) (6 \theta + 1) 
     \right]
     \Pi_{\widetilde{\mathbf{WCP}}^2_{3,1,2}[6]} \, ,
\\[1ex]
0 &= \left[ 
            \theta^2 - 4 e^{-t} (4 \theta + 3) (4 \theta + 1) 
     \right]
     \Pi_{\widetilde{\mathbf{WCP}}^2_{2,1,1}[4]} \, ,  
\end{align}
respectively.
For rows 3 and 4 of Table \ref{Table-CP^3[4]}, we obtain
\begin{align}
0 &= \Bigl\{ 
            137,781 
            \left[
                   {\textstyle \prod_{i=1}^7} (8 \theta - i) 
            \right]
            \left[
                   {\textstyle \prod_{j=1}^6} (7 \theta - j) 
            \right]
            (3 \theta - 2) (3 \theta - 1) \theta^3
     \Bigr.
\nonumber
\\
  &\phantom{= \left\{ \right. }
   + 962,990,300,932 \, e^{-t} 
     (36 \theta + 35) (36 \theta + 34) (36 \theta + 31) (36 \theta + 29) (36 \theta + 26) 
\nonumber
\\[1ex]
  &\phantom{= \left\{ \right. +}
     (36 \theta + 25) (36 \theta + 23) (36 \theta + 22) (36 \theta + 19) (36 \theta + 17) 
     (36 \theta + 14) (36 \theta + 13) 
\nonumber
\\[1ex]
  &\phantom{= \left\{ \right. +}
   \Bigl.
    (36 \theta + 11) (36 \theta + 10) (36 \theta + 7) (36 \theta + 5) (36 \theta + 2) 
    (36 \theta + 1)  
   \Bigr\}                                 
    \Pi_{\widetilde{\mathbf{WCP}}^3_{12,8,7,9}[36]} \, ,
\\[1ex]
0 &= \Bigl\{ 
            280
            \left[
                   {\textstyle \prod_{i=1}^6} (7 \theta - i) 
            \right]
            (6 \theta - 5) (6 \theta - 1)
            \left[
                   {\textstyle \prod_{j=1}^4} (5 \theta - j) 
            \right]
            (3 \theta - 2) (3 \theta - 1) (2 \theta - 1) \theta^3
     \Bigr.
\nonumber
\\
  &\phantom{= \left\{ \right. }
   + 2187 \, e^{-t} 
     (27 \theta + 26) (27 \theta + 25) (27 \theta + 23) (27 \theta + 22) (27 \theta + 20)
     (27 \theta + 19)
\nonumber
\\[1ex]
  &\phantom{= \left\{ \right. +}      
     (27 \theta + 17) (27 \theta + 16) (27 \theta + 14) (27 \theta + 13)
     (27 \theta + 11) (27 \theta + 10) (27 \theta + 8)
\nonumber
\\[1ex]
  &\phantom{= \left\{ \right. +}
   \Bigl.
    (27 \theta + 7) (27 \theta + 5) (27 \theta + 4) (27 \theta + 2) 
    (27 \theta + 1)  
   \Bigr\}                                 
    \Pi_{\widetilde{\mathbf{WCP}}^3_{9,6,7,5}[27]} \, ,  
\end{align}
respectively.
Finally, for rows 6 and 9 of Table \ref{Table-CP^4[5]}, we obtain 
\begin{align}
0 &= \Bigl\{ 
            9 
            \left[
                   {\textstyle \prod_{i=1}^4} (5 \theta - i) 
            \right]
            (3 \theta - 2)^2 (3 \theta - 1)^2 \theta^4
     \Bigr.
\nonumber
\\
  &\phantom{= \left\{ \right. }
   + 12,800 \, e^{-t} 
     (20 \theta + 19) (20 \theta + 18) (20 \theta + 17) (20 \theta + 14) (36 \theta + 13)
     (20 \theta + 11)
\nonumber
\\[1ex]
  &\phantom{= \left\{ \right. +}      
     (20 \theta + 9) (20 \theta + 7) (20 \theta + 6) (20 \theta + 3)
     (20 \theta + 2) (20 \theta + 1)
   \Bigl.
   \Bigr\}                                 
    \Pi_{\widetilde{\mathbf{WCP}}^4_{5,3,5,3,4}[20]} \, ,
\\[1ex]
0 &= \Bigl\{ 
             3 
             (4 \theta - 3)^2 (4 \theta - 1)^2
             (3 \theta - 2) (3 \theta - 1) (2 \theta - 1)^2 \theta^4
     \Bigr.
\nonumber
\\
  &\phantom{= \left\{ \right. }
   + 1000 \, e^{-t} 
     (20 \theta + 19) (20 \theta + 18) (20 \theta + 17) (20 \theta + 14) (20 \theta + 13)
     (20 \theta + 11)
\nonumber
\\[1ex]
  &\phantom{= \left\{ \right. +}
   \Bigl.      
     (20 \theta + 9) (20 \theta + 7) (20 \theta + 6)
     (20 \theta + 3)
     (20 \theta + 2) (20 \theta + 1)  
   \Bigr\}                                 
    \Pi_{\widetilde{\mathbf{WCP}}^4_{4,5,3,4,4}[20]} \, ,  
\end{align}
respectively.
Note that the Picard-Fuchs operators appearing in each of the above pairs of 
Picard-Fuchs equations are different from each other.
This indicates that they correspond to different K\"{ahler} structures for 
$ \widetilde{\mathcal{M}}_{\mathcal{W}} $.

\section{\label{Discussion}Discussion}

In Section \ref{CYPF}, we found that the technique suggested by Hori and Vafa 
\cite{HoriVafa:Mirror} for determining the Picard-Fuchs equations satisfied by 
$ \Pi_{\widetilde{M}_G} $ when the Calabi-Yau condition (\ref{c_1(M_G)=0}) holds yields results which agree with those obtained in 
\cite{HosonoKlemmTheisenYau:Mirror,LianYau:Mirror} working in the Batyrev-Borisov 
\cite{Batyrev:Dual,Borisov:Towards} framework.
An advantage of the Hori-Vafa formalism is that it provides an explicit expression for
$ \Pi_{\widetilde{M}_G} $.
The case in which the Calabi-Yau condition is replaced by (\ref{nonnegative-c_1(M_G)}) can be treated by making a minor modification to (\ref{GKZ-mirror-period-N(mu,t)}) and
(\ref{GKZ-mirror-period-N(T)}).

In Tables \ref{Table-CP^2[3]}, \ref{Table-CP^3[4]}, and \ref{Table-CP^4[5]}, the left columns correspond to different complex structures for $ \mathcal{M}_{\mathcal{W}} $ whereas the right columns correspond to different K\"{a}hler structures for 
$ \widetilde{\mathcal{M}}_{\mathcal{W}} $.
By choosing the change of variables (\ref{change-of-variables}) appropriately, the Hori-Vafa formalism allows any of the K\"{a}hler structures for $ \widetilde{\mathcal{M}}_{\mathcal{W}} $ to be obtained in the limit $ t \rightarrow - \infty $.
However, the Hori-Vafa formalism provides no prescription for associating a particular complex structure for $ \mathcal{M}_{\mathcal{W}} $ with a particular K\"{a}hler structure for 
$ \widetilde{\mathcal{M}}_{\mathcal{W}} $.
For this, we have made use of the Chiodo-Ruan \cite{ChiodoRuan:LG/CY} geometric interpretation of Berglund-H\"{u}bsch-Krawitz duality.
We have probed some of the resulting mirror K\"{a}hler structures by determining corresponding Picard-Fuchs equations.

Generally speaking, when $ \langle J_{\mathcal{W}} \rangle \subset \mathrm{SL}_{\mathcal{W}} $, this corresponds to
\begin{equation*}
\left\{ \mathcal{W}(\phi_1,\ldots,\phi_m) = 0 \right\}
   \in \mathbf{WCP}^{m-1}_{n_{\phi_1},\ldots,n_{\phi_m}}[s] \, ,
\qquad    
   \frac{1}{s} \sum_{i=1}^m n_{\phi_i} = \kappa \in \mathbf{Z}_{>0} \, .  
\end{equation*}
The case $ \kappa = 1 $ is the Calabi-Yau condition.
The case $ \kappa > 1 $ can be described in terms of the special class of Fano varieties discussed in 
\cite{Schimmrigk:Critical,Schimmrigk:Mirror,CandelasDerrickParkes:Generalized}.
Alternatively, for the $ \kappa > 1 $ case, the Landau-Ginzburg orbifold 
$ \mathcal{W} / \langle J_{\mathcal{W}} \rangle $ can be given a geometrical interpretation as a nonlinear sigma model on a \emph{super} Calabi-Yau using the proposed correspondence of Sethi \cite{Sethi}; see also
\cite{GaravusoKreuzerNoll:Fano,GaravusoKatzarkovKreuzerNoll,GaravusoKatzarkovNoll}. 
It was noted in \cite{GaravusoKreuzerNoll:Fano} that this super Calabi-Yau should be equivalent (in the sense of \cite{Schwarz:Sigma}) to a Calabi-Yau complete intersection when the Newton polytope associated with $ \mathcal{W} $ admits a nef partition.
The Batyrev-Borisov construction yields a mirror for this Calabi-Yau complete intersection.
Borisov \cite{Borisov:Berglund-Huebsch} has suggested a way to unify the Batyrev-Borisov and Berglund-H\"{u}bsch constructions.

The above comments illustrate the overlapping nature of various mirror symmetry formalisms.
While this paper has helped elucidate some of these overlaps, a complete mirror symmetry ``Venn diagram" has still not been achieved.

\section*{Acknowledgements} 

C.D. thanks The Institute of Mathematical Sciences at The Chinese University of Hong Kong for their hospitality during the final stages of this project.  
R.G. thanks Y. Ruan for useful discussions regarding admissible groups. 
R.G. received financial support from NSERC.  


\end{document}